\documentclass[journal]{IEEEtran}
\usepackage{algorithm2e}
\usepackage{cite}
\usepackage{graphicx}
\usepackage{psfrag}
\usepackage{subfigure}
\usepackage{url}
\usepackage{amsmath}
\usepackage{array}
\usepackage{amssymb}
\usepackage{amsfonts}
\newtheorem{proposition}{Proposition}
\newtheorem{lemma}{Lemma}
\newtheorem{observation}{Observation}

\newtheorem{remark}{Remark}
\newtheorem{definition}{Definition}

\newtheorem{example}{Example}

\title{Single-generation Network Coding for Networks with Delay}
\author{
\authorblockN{K.~Prasad and B.~Sundar Rajan}
}	
\date{\today}
\begin{document}
\maketitle
\thispagestyle{empty}
\begin{abstract}
A single-source network is said to be \textit{memory-free} if all of the internal nodes (those except the source and the sinks) do not employ memory but merely send linear combinations of the incoming symbols (received at their incoming edges) on their outgoing edges. Memory-free networks with delay using network coding are forced to do inter-generation network coding, as a result of which the problem of some or all sinks requiring a large amount of memory for decoding is faced. In this work, we address this problem by utilizing memory elements at the internal nodes of the network also, which results in the reduction of the number of memory elements used at the sinks. We give an algorithm which employs memory at the nodes to achieve single-generation network coding. For fixed latency, our algorithm reduces the total number of memory elements used in the network to achieve single-generation network coding. We also discuss the advantages of employing single-generation network coding together with convolutional network-error correction codes (CNECCs) for networks with unit-delay and illustrate the performance gain of CNECCs by using memory at the intermediate nodes using simulations on an example network under a probabilistic network error model. 
\end{abstract}	
\section{Introduction}
Network coding was introduced in \cite{ACLY} as a means of achieving maximum rate of transmission in wireline networks. An algebraic formulation of network coding was discussed in \cite{KoM} for both instantaneous networks and networks with delays. Convolutional network-error correcting codes(CNECCs) were	introduced for acyclic instantaneous networks in \cite{PrR} and for unit-delay, memory-free networks in \cite{PrR2}.

In this work, we consider acyclic, single-source networks with delays which have a multicast network code in place. The set of all code symbols generated at the source at any particular time instant is called a \textit{generation}. In unit-delay, memory-free networks, the nodes of the network may receive information of different generations on their incoming edges at every time instant and therefore network coding across generations (\textit{inter-generation}) is unavoidable in general. However, the sinks have to employ memory to decode the symbols. If memory is utilized in the internal nodes also, such inter-generation network coding can be avoided thus making the decoding simpler. 

We define a  \textit{single-generation network code} as a network code where all the symbols received at all the sinks are linear combinations of the symbols belonging to the same generation. In \cite{WZY}, the technique of adding memory at the nodes to achieve single-generation network coding was discussed. However this was done only on a per-node basis without considering the entire topology or the network code of the network. On the other hand, we consider the entire network topology and the network code, which govern the addition of memory elements at the nodes and the way in which they are rearranged across the network to reduce the overall memory usage in the network. 

The organization and contributions of this work are as follows 
\begin{itemize}
\item After briefly discussing the network setup and the network code for an acyclic network with delays and memory (Section \ref{sec2}), we introduce different methods of adding memory at a node and analyze how each of them affect the local and global encoding kernels of the network code (Section \ref{sec3}).
\item We also present different memory reduction and distribution techniques (Section \ref{sec4}). 
\item We propose an algorithm which uses the memory at the nodes to achieve single-generation network coding while reducing the overall memory usage in the network (Section \ref{sec5}).
\item We discuss the advantages of employing memory at the intermediate nodes in tandem with CNECCs in terms of their encoding/decoding (Section \ref{sec6}).
\item We illustrate the the performance benefits by using memory for CNECCs for unit-delay networks using simulations on an example unit-delay network under a probabilistic error setting (Section \ref{sec7}). 
\end{itemize}
\section{Networks with delay and memory}
\label{sec2}
The model for acyclic networks with delays considered in this paper is as in \cite{KoM}. An acyclic network can be represented as an acyclic directed multi-graph (a graph that can have parallel edges between nodes) ${\cal G}$ = ($\cal V,\cal E$) where $\cal V$ is the set of all vertices and $\cal E$ is the set of all edges in the network. 

We assume that every edge in the directed multi-graph representing the network has unit \emph{capacity} (can carry utmost one symbol from $\mathbb{F}_q$, the field with $q$ elements). Network links with capacities greater than unit are modeled as parallel edges. The network has delays, i.e, every edge in the directed graph representing the input has a unit delay associated with it, represented by the parameter~$z$. Such networks are known as \textit{unit-delay networks}. Those network links with delays greater than unit are modeled as serially concatenated edges in the directed multi-graph. We assume a single-source node $s\in\cal V$ and a set of sinks $\cal T$. Let $n_{_T}$ be the unicast capacity for a sink node  $T\in{\cal T}$ i.e the maximum number of edge-disjoint paths from $s$ to $T$. Then 
\[
n_{min} = \min_{T\in{\cal T}}n_{_T}
\]
is the max-flow min-cut capacity of the multicast connection.

%

\subsection{Network code for unit-delay, memory-free networks}	
We follow  \cite{KoM} in describing the network code. For each node $v\in {\cal V}$, let the set of all incoming edges be denoted by $\Gamma_I(v)$. Then $|\Gamma_I(v)|=\delta_I(v)$ is the in-degree of $v$. Similarly the set of all outgoing edges is defined by $\Gamma_O(v)$, and the out-degree of the node $v$ is given by $|\Gamma_O(v)|=\delta_O(v)$. 

For any $e \in {\cal E}$ and $v \in {\cal V}$, let $head(e)=v$, if $v$ is such that $e \in \Gamma_I(v)$. Similarly, let $tail(e)=v$, if $v$ is such that $e \in \Gamma_O(v)$. We will assume an ancestral ordering on ${\cal V}$ 
and ${\cal E}$ 
of the acyclic graph of the unit-delay, memory-free network.

The network code can be defined by the local kernel matrices of size $\delta_I(v)\times\delta_O(v)$ for each node $v\in{\cal V}$ with entries from $\mathbb{F}_q$. The global encoding kernels for each edge can be recursively calculated from these local kernels.

The network transfer matrix, which governs the input-output relationship in the network, is defined as given in \cite{KoM} for an $n$-dimensional ($n\leq n_{min}$) network code. Towards this end, the matrices $A$,$K$,and $B^T$(for every sink $T\in {\cal T}$) are defined as follows.

The entries of the $n \times |{\cal E}|$ matrix $A$ are defined as
\[
A_{i,j}=\left\{
\begin{array}{cc}
\alpha_{i,e_j} & \text{   if } e_j \in \Gamma_{O}(s)\\
0  & \text{ otherwise}
\end{array}
\right. 
\]
where $\alpha_{i,e_j} \in \mathbb{F}_q$ is the local encoding kernel coefficient at the source coupling input $i$ with edge $e_j \in \Gamma_O(s)$.

The $(i,j)^{th}$ entry of the $|{\cal E}| \times |{\cal E}|$ matrix $K$ is $K_{e_i,e_j}\in \mathbb{F}_q$ which is the local kernel coefficient between $e_i$ and $e_j$  at the node $head(e_i) = tail(e_j)$ (if such a node exists), and zero if $head(e_i) \neq tail(e_j)$.

For every sink $T \in {\cal T}$, the entries of the $|{\cal E}| \times n$ matrix $B^T$ are defined as  
\[
B^T_{i,j}=\left\{
\begin{array}{cc}
\epsilon_{e_j,i} & \text{   if } e_j \in \Gamma_{I}(T)\\
0  & \text{ otherwise} 
\end{array} 
\right. 
 \]
where all $\epsilon_{e_j,i} \in \mathbb{F}_q$.

For unit-delay, memory-free networks, we have 
\begin{eqnarray*}
F(z): = (I-zK)^{-1} 
\end{eqnarray*}
where $I$ is the $|{\cal E}| \times |{\cal E}|$ identity matrix.  Now we have the following definition.
\begin{definition}[\cite{KoM}]
\label{nettransmatrix}
\textit{The network transfer matrix}, $M_{T}(z)$, corresponding to a sink node ${T} \in \cal T$ for a $n$-dimensional network code, is a full rank (over the field of rationals $\mathbb{F}_q(z)$) $n \times n$ matrix defined as 
\[
M_{T}(z):=AF(z)B^{T}=AF_{T}(z).
\] 
\end{definition} 	

With an $n$-dimensional network code, the input and the output of the network are $n$-tuples of elements from $\mathbb{F}_q[[z]]$, the formal power series ring over $\mathbb{F}_q.$ Definition \ref{nettransmatrix} implies that if $\boldsymbol{x}(z) \in \mathbb{F}_q^n[[z]]$ is the input to the unit-delay, memory-free network, then at any particular sink $T \in \cal T$, we have the output, $\boldsymbol{y}(z) \in \mathbb{F}_q^n[[z]]$, to be 
\[
\boldsymbol{y}(z) = \boldsymbol{x}(z)M_T(z).
\]
\subsection{Network code for networks with delay and memory}
\begin{figure}[htbp]
\centering
\includegraphics[totalheight=4.4in,width=2.8in]{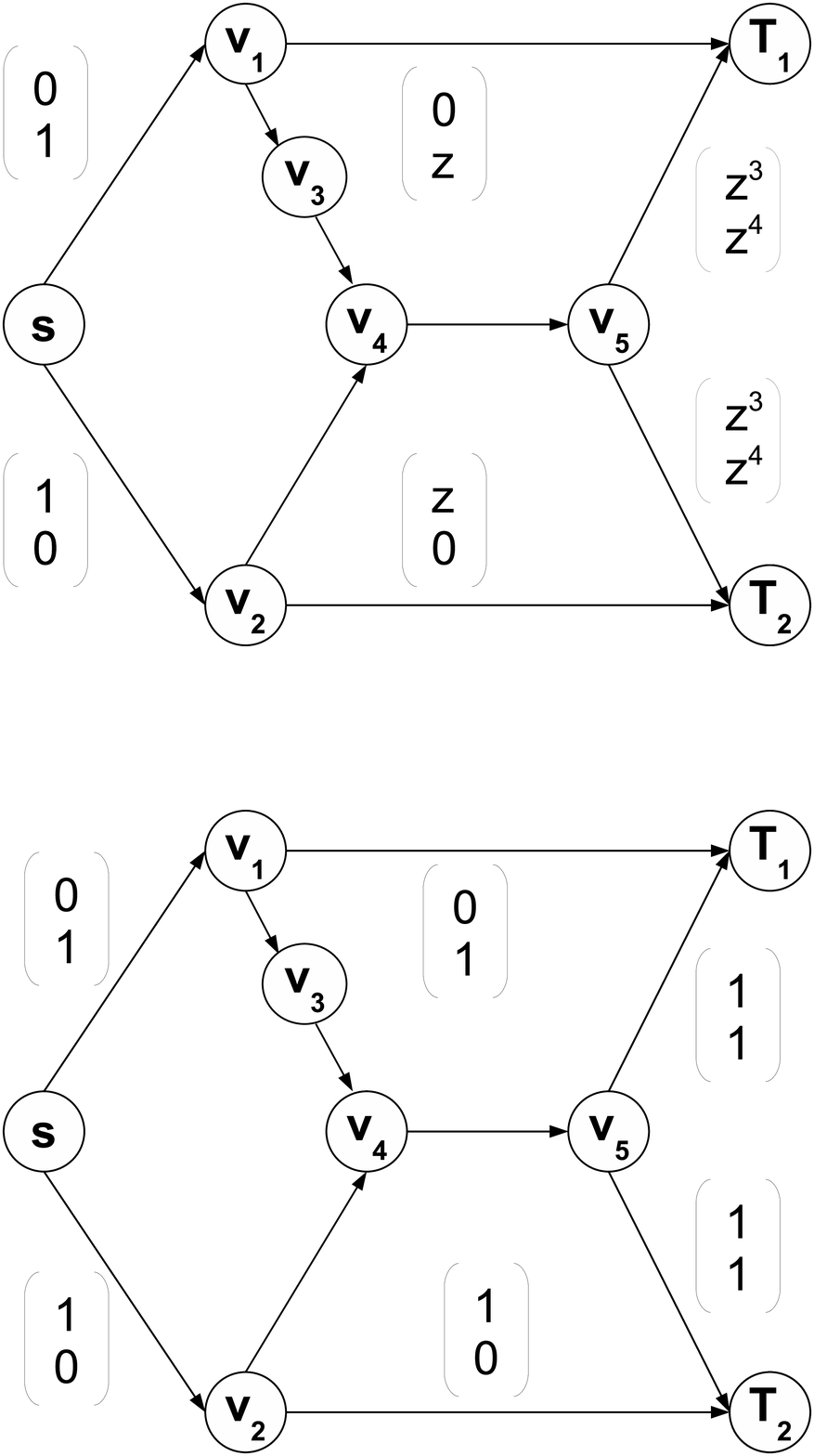}
\caption{The figure corresponding to Example \ref{ex} (A unit-delay network and its instantaneous counterpart). }	
\label{fig:instantcounter}	
\end{figure}

We define the \textit{instantaneous counterpart} of a unit-delay network as follows.
\begin{definition}
Given a unit-delay network ${\cal G}({\cal V},{\cal E})$, the network obtained from ${\cal G}$ (having the same node set ${\cal V}$ and the same edge set ${\cal E}$) by removing the delays associated with the edges is defined as the \textit{instantaneous counterpart} of ${\cal G}({\cal V},{\cal E}).$ 
\end{definition}
\begin{example}
\label{ex}
Fig. \ref{fig:instantcounter} illustrates an example. A modified butterfly unit-delay network (top) and its instantaneous counterpart (bottom) are shown. The global kernels of the incoming edges to the sinks $T_1$ and $T_2$ corresponding to a $2$ dimensional network code are indicated for both networks.
\end{example}

Let ${\cal G}_m({\cal V},{\cal E})$ be a single-source, acyclic network with every edge of the network having some delay (a positive integer) and with memory elements at the nodes available for usage. If none of the memory elements at the nodes are used, then we can model ${\cal G}_m$ as a unit-delay, memory-free network ${\cal G}_u.$ Let ${\cal G}_{inst}$ be the instantaneous counterpart of ${\cal G}_u.$ The following lemma ensures the equivalence of a network code between ${\cal G}_{inst}$ and ${\cal G}_u.$
\begin{lemma}[\cite{PrR2} ]
\label{lemma1}
Let ${\cal G}'({\cal V},{\cal E})$ be a single-source acyclic, unit-delay, memory-free network, and ${\cal G}'_{inst}$ be the instantaneous counterpart of ${\cal G}'.$ Let ${\cal N}$ be the set of all $\delta_I(v) \times \delta_O(v)$ matrices $\forall$ $v \in \cal V$, i.e, the set of local encoding kernel matrices at each node, describing an $m$-dimensional network code (over $\mathbb{F}_q$) for ${\cal G}'_{inst}$ ($m \leq$ min-cut of the source-sink connections in ${\cal G}'_{inst}$). Then the network code described by $\cal N$ continues to be an $m$-dimensional network code (over $\mathbb{F}_q(z)$) for the unit-delay, memory-free network ${\cal G}'.$
\end{lemma}

If the nodes use memory elements such that inter-generation network coding is prevented at any particular node of the network, then this leads to single-generation network coding in the network. 

In Section \ref{sec5} we give an algorithm which uses memory elements at the nodes to achieve single-generation network coding, i.e, the network transfer matrix $M_T(z)$ of every sink $T\in \cal T$ in the  in ${\cal G}_m$ becomes 
\begin{equation}
\label{eqn28}
M_T(z)=z^{L_T}M_T
\end{equation}
where $L_T$ is some positive integer and $M_T$ is the network transfer matrix of the sink $T$ in ${\cal G}_{inst}.$ Clearly, if $M_T$ is full rank (over $\mathbb{F}_q$), so is $M_T(z)$ (over $\mathbb{F}_q(z)$).
\section{Memory additions at a node}
\label{sec3}
For the source node $s$, let $\tilde{\Gamma}_I(s)$ denote the set of $n$ virtual incoming edges which denote the $n$ inputs. The global kernels of these edges are therefore the columns of an $n\times n$ identity matrix over $\mathbb{F}_q$, the field over which the network code is defined. For every non-source node $v \in \cal V$, let $\tilde{\Gamma}_I(v)=\phi.$ For a sink $T \in \cal T$, let $\tilde{\Gamma}_O(T)$ denote $n$ virtual outgoing edges denoting the $n$ outputs at sink $T.$ The global kernels of these edges are the columns of the network transfer matrix $M_T(z).$ For every non-sink node $v \in \cal V$, let $\tilde{\Gamma}_O(v)=\phi.$ We then define the set $\tilde{{\cal E}}$ as 
\[
\tilde{{\cal E}}:={\cal E}\cup\tilde{\Gamma}_I(s)\cup\left(\bigcup_{T\in{\cal T}}\tilde{\Gamma}_O(T)\right)
\]

The ancestral ordering on $\cal E$ can then be extended to an ancestral ordering on $\tilde{\cal E}$. 

For any $e_i, e_j \in \tilde{{\cal E}}$ such that $head(e_i)=tail(e_j)=v \in {\cal V}$, with memory being used at $v$, the local kernel $A_{e_i,e_j}$ (the kernel coefficient between $e_i\in \tilde{\Gamma}_I(s)$ and $e_j\in \Gamma_O(s)$ with $s=v$), $K_{e_i,e_j}$ or $B^{v}_{e_i,e_j}$ (the kernel coefficient between $e_i$ and $e_j\in \tilde{\Gamma}_O(v)$ for some sink node $v$) can have elements from $\mathbb{F}_q(z).$ We show in Section \ref{sec5} that using the memory elements at the nodes according to Subsection \ref{pairmem} and Subsection \ref{outmem} is sufficient to guarantee single-generation network coding at each node and therefore in the given network.
\subsection{Adding memory at a node for a pair of an incoming and an outgoing edge}
\label{pairmem}
For any $e_i, e_j \in \tilde{{\cal E}}$ such that $head(e_i)=tail(e_j)=v \in {\cal V}$, we define $M_{e_i,e_j}$ as the number of memory elements utilized at the node $v$ to delay the symbols coming from the incoming edge $e_i$ (before any network coding is performed at node $v$ on the symbols from $e_i$) such that the local kernel between $e_i$ and $e_j$ is modified in one of the following ways 
\begin{align}
\label{eqn31}
A_{e_i,e_j} &\longmapsto z^{M_{e_i,e_j}}A_{e_i,e_j}~~~\text{if}~e_i \in  \tilde{\Gamma}_I(s), e_j \in {\cal E} \\
\label{eqn21}
K_{e_i,e_j} &\longmapsto z^{M_{e_i,e_j}}K_{e_i,e_j}~~~\text{if}~e_i, e_j \in {\cal E} \\
\label{eqn32}
B^{v}_{e_i,e_j} &\longmapsto z^{M_{e_i,e_j}}B^v_{e_i,e_j}~~~\text{if}~e_i \in {\cal E}, e_j \in  \tilde{\Gamma}_O(v) 
\end{align}
while none of the other local kernels are changed. The matrix $F(z)=(I-zK)^{-1}$ is also correspondingly modified.
\subsection{Adding memory at a node for an outgoing edge}	
\label{outmem}
\begin{figure}[htbp]
\centering
\includegraphics[totalheight=2.4in,width=3.6in]{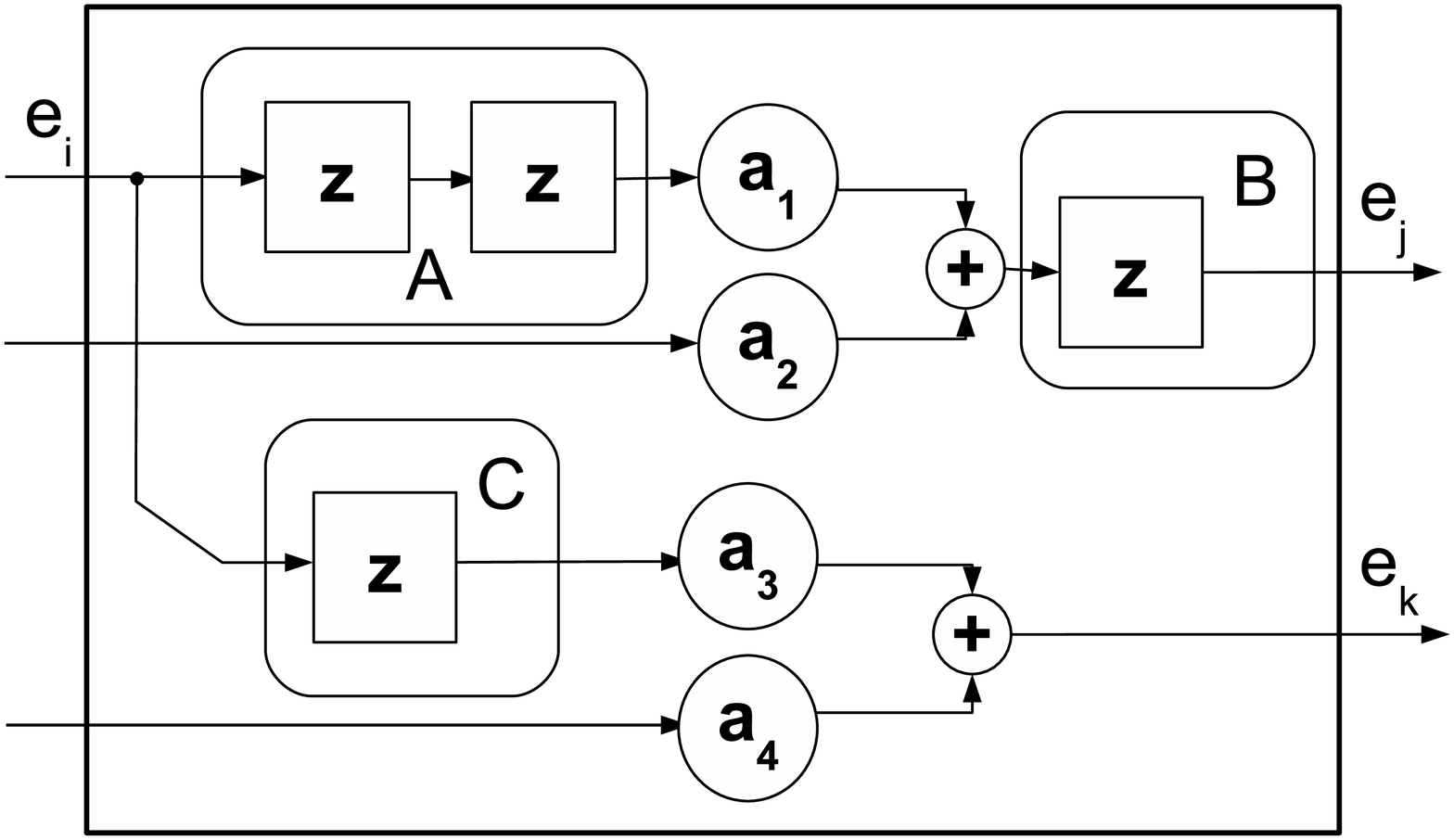}
\caption{The figure corresponding to Example \ref{ex1} (Adding memory at a node).}	
\label{fig:memoryadditions}	
\end{figure}

For $e_j \in \Gamma_O(v)\cup\tilde{\Gamma}_O(v)$, we define $M_{e_j,tail(e_j)}$ as the number of memory elements added at node $v$ to delay the symbols going into the edge $e_j$ after performing network coding at $v.$ In such a case, the elements of the matrix $K$ (or of the matrix or $A$, or $B^v$) are modified according to the following rule.
\begin{align}
\label{eqn33}
A_{e_i,e_j}& \longmapsto z^{M_{e_j,tail(e_j)}}A_{e_i,e_j}&\forall& e_i \in \Gamma_{I,e_j}(v),~\text{if}~v=s\\
\label{eqn22}
K_{e_i,e_j}& \longmapsto z^{M_{e_j,tail(e_j)}}K_{e_i,e_j}&\forall& e_i \in \Gamma_{I,e_j}(v),\\
\nonumber &&&~\text{if}~v \neq s, e_j \in \Gamma_O(v)\\
\label{eqn34}
B^v_{e_i,e_j}& \longmapsto z^{M_{e_j,tail(e_j)}}B^v_{e_i,e_j}&\forall& e_i \in \Gamma_{I,e_j}(v),\text{if}~e_j \in \tilde{\Gamma}_O(v)
\end{align}
where the set $\Gamma_{I,e_j}(v) \subseteq \Gamma_{I}(v)\cup\tilde{\Gamma}_{I}(v)$ is defined as in the top of the next page.
\begin{figure*}
\begin{equation}
\label{eqn27}
\Gamma_{I,e_j}(v):=\left\{e_i\in\Gamma_{I}(v)~|~K_{e_i,e_j} \neq 0\right\}\bigcup\left\{e_i\in\tilde{\Gamma}_{I}(v)~|~A_{e_i,e_j} \neq 0\right\}.
\end{equation}
\hrule
\end{figure*}
The elements of the matrix $F(z)$ are also correspondingly modified.
\begin{example}
\label{ex1}
Fig \ref{fig:memoryadditions} illustrates an example of the memory additions at a node. The memory elements indicated inside the box labeled `A' are added at the node for the pair of edges $e_i$ and $e_j$ thereby delaying the symbols on $e_i$ before network coding at the node, i.e, $M_{e_i,e_j}=2.$ Similarly the memory element indicated by `C' is added for the pair of edges $e_i$ and $e_k$, i.e, $M_{e_i,e_k}=1.$ The memory element indicated by `B' is added for the outgoing edge $e_j$ after network coding, i.e, $M_{e_j,tail(e_j)}=1.$
\end{example}
\section{Memory reduction and distribution techniques}
\label{sec4}
In this section, we look at techniques to reduce the memory used at the nodes of the network and the overall memory used in the network and also to obtain a fairly uniform memory usage distribution throughout the network. 

We define the maximum number of memory elements added to delay the symbols coming from an edge $e_i \in \tilde{\cal E}$ into node $head(e_i)=v$ as 
\begin{eqnarray}
\label{Memax}
M_{e_i,head(e_i),max}: = \max_{e_j \in \Gamma_{O,e_i}(v)}Me_i,e_j 
\end{eqnarray}
where $\Gamma_{O,e_i}(v)$ is defined as shown at the top of the next page. 
\begin{figure*}
\begin{align}
\label{eqn26}
\Gamma_{O,e_i}(v):=\left\{e_j\in\Gamma_{O}(v)~|~K_{e_i,e_j} \neq 0\right\}\bigcup\left\{e_j\in\tilde{\Gamma}_{O}(v)~|~B^v_{e_i,e_j} \neq 0\right\}
\end{align}
\hrule
\end{figure*}
We define the total number of memory elements used at node $v$ as
\[
M_v = \sum_{e_i \in \Gamma_I(v)\cup\tilde{\Gamma}_I(v)} M_{e_i,head(e_i),max} + \sum_{e_j \in \Gamma_O(v)\cup\tilde{\Gamma}_O(v)}M_{e_j,tail(e_j)}.
\]
\subsection{Memory reduction in a single node}
\label{sec4a}
\label{nodememreduction}
Consider a node $v \in {\cal V}$ in which memory elements have been added to delay symbols coming from an edge $e_i \in \Gamma_I(v)\cup\tilde{\Gamma}_I(v).$ 

\begin{figure}[htbp]
\centering
\includegraphics[totalheight=4in,width=3in]{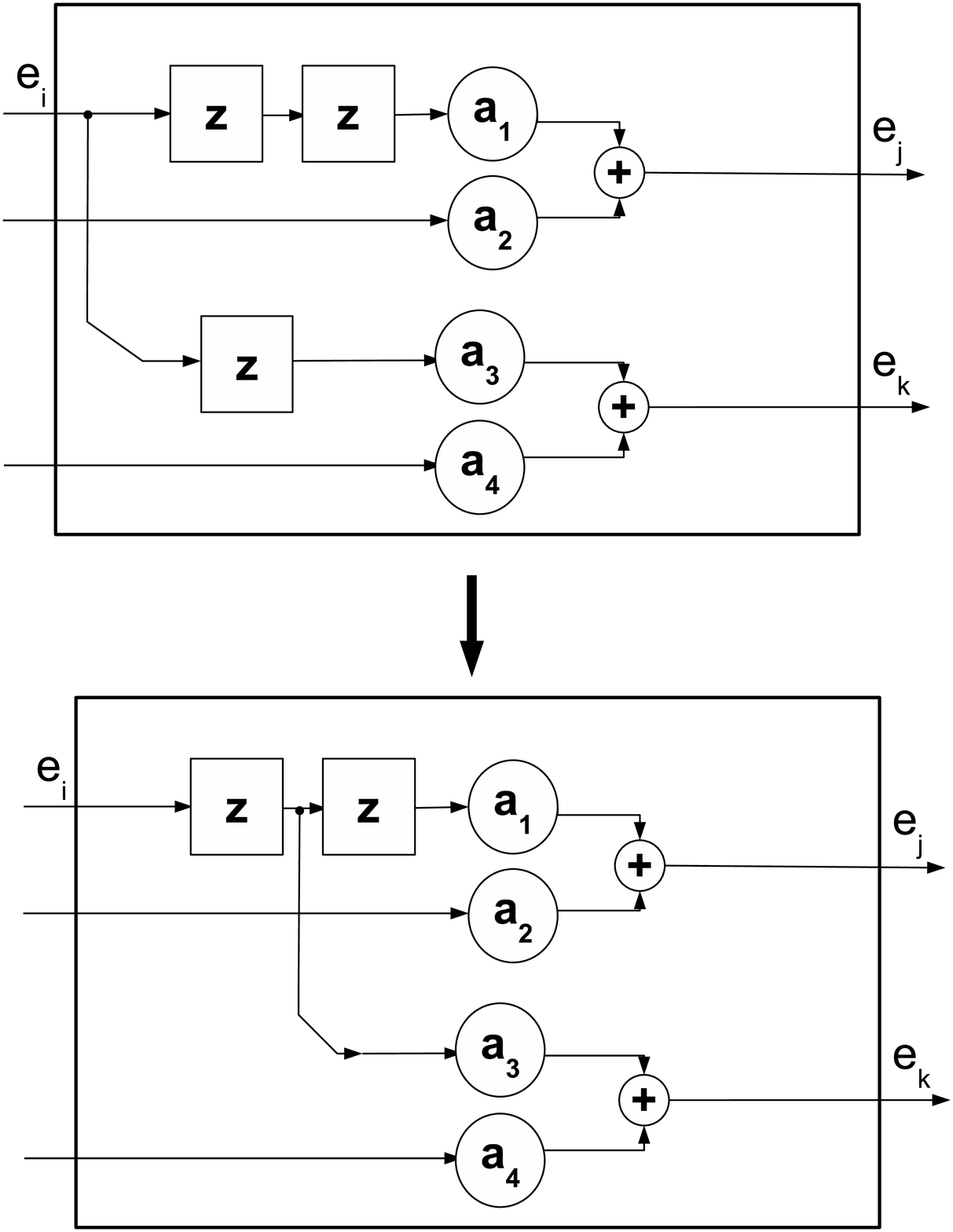}
\caption{The figure corresponding to Example \ref{ex0} (Memory reduction at a node).}	
\label{fig:memnodereduction}	
\end{figure}

Then, retaining the $M_{e_i,head(e_i),max}$(as defined in (\ref{Memax})) memory elements, all other memory elements placed on $e_i$ can be removed without any change in any local or global kernels by tapping symbols from the $M_{e_i,head(e_i),max}$ memory elements wherever necessary. Doing this for every incoming edge of $v$ is equivalent to obtaining a minimal encoder (one with minimum number of memory elements) of the transfer function (input-output relationship) at node $v.$ 
\begin{example}
\label{ex0}
Fig. \ref{fig:memnodereduction} illustrates a particular example of such a reduction. The figure on the top (all $a_i \in \mathbb{F}_q$) represents a node $v$ before memory reduction with $M_v=3$, while the figure on the bottom is the same node after memory reduction with $M_v=2.$
\end{example}
\subsection{Memory reduction between nodes}

For a set of edges ${\cal E}'\subseteq \tilde{\cal E}$, let ${\cal V}_{{\cal E}'}$ be the set of all nodes defined as follows
\begin{equation}
\label{eqn40}
{\cal V}_{{\cal E}'} = \left\{ head(e_j)~|~e_j\in{\cal E}' \right\}
\end{equation}

We now define $M_{e_i,head(e_i),min}$ and $M_{{\cal E}'}$ as follows. 
\begin{eqnarray}
\label{eqn38}
M_{e_i,head(e_i),min}: = \min_{e_j \in \Gamma_{O,e_i}(v)}Me_i,e_j\\
\label{eqn39}
M_{{\cal E}'}:=\min_{e_j \in {\cal E}'}M_{e_j,head(e_j),min}
\end{eqnarray}
where $\Gamma_{O,e_i}(v)$ is as defined in (\ref{eqn26}). 

For a node $v\in\cal V$, we define the set of \textit{adjacent nodes} of $v$ as the set of nodes 
\[
{\cal E}_v:=\left\{v'~|~v'=head(e_j)~\forall e_j\in\Gamma_O(v)\right\}.
\]
\subsubsection{Memory reduction between adjacent nodes}
\label{distributedmemreduction1}
For a node $v \in \cal V$, and for some $\Gamma_{O}'(v) \subseteq \Gamma_{O}(v)\cup\tilde{\Gamma}_{O}(v)$, let $\Gamma_{I}'(v) \subseteq \Gamma_{I}(v)\cup\tilde{\Gamma}_I(v)$ be defined as 
\[
\Gamma_{I}'(v) = \bigcup_{e_j \in \Gamma_{O}'(v)}\Gamma_{I,e_j}(v).
\]
where $\Gamma_{I,e_j}(v)$ is as in (\ref{eqn27}), i.e, the global kernels of the edges in $e_j \in \Gamma_{O}'(v)$ are linear combinations of the global kernels of the edges in $\Gamma_{I}'(v)$ only and none else. Also let $M_{\Gamma_{O}'(v)}$ and the set ${\cal V}_{\Gamma_{O}'(v)} \subseteq {\cal E}_v$ of nodes be defined for the set of edges $\Gamma_{O}'(v)$ as in (\ref{eqn39}) and (\ref{eqn40}) respectively. 

We define the term $M_{e_i,\Gamma_{O}'(v)}$ as
\begin{equation}
\label{eqn43}
M_{e_i,\Gamma_{O}'(v)}=\max\left\{0,M_{\Gamma_{O}'(v)}-M_{e_i,head(e_i),max}\right\}
\end{equation}

Then, if the condition is satisfied,
\begin{equation}
\label{eqn29}
\sum_{e_i \in \Gamma_{I}'(v)}M_{e_i,\Gamma_{O}'(v)} \leq M_{\Gamma_{O}'(v)}|\Gamma_{O}'(v)| 
\end{equation}
then all of the $|\Gamma_{O}'(v)|M_{\Gamma_{O}'(v)}$ used at the nodes ${\cal V}_{\Gamma_{O}'(v)}$ (to delay symbols coming from the edges $e_j \in \Gamma_{O}'(v)$) can be `absorbed' into node $v$ by removing all these memory elements and adding $M_{e_i,\Gamma_{O}'(v)}$ memory elements at node $v$ for every $e_i \in \Gamma_{I}'(v)$ (and thereby used for delaying the symbols coming from every $e_i \in \Gamma_{I}'(v)$), without using any additional memory and without changing the global kernels of any outgoing edge of any node in ${\cal V}_{\Gamma_{O}'(v)}$. 

\begin{figure}[htbp]
\centering
\includegraphics[totalheight=4in,width=3.1in]{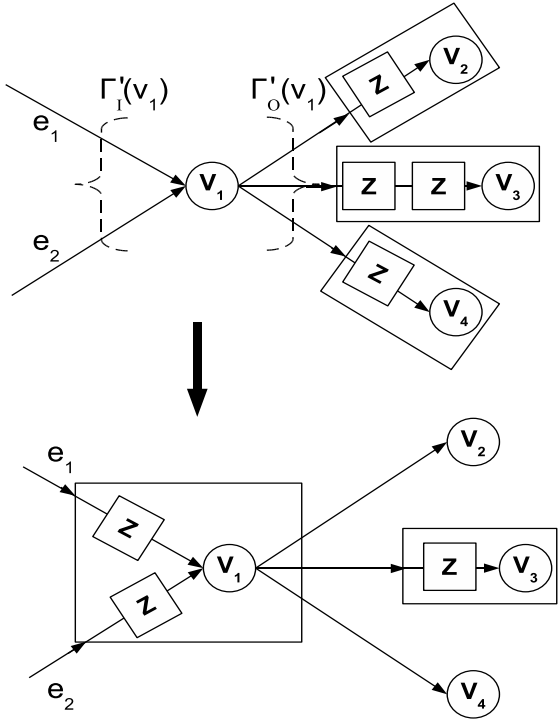}
\caption{The figure corresponding to Example \ref{ex2} (Memory reduction between adjacent nodes).} 
\label{fig:memnodereductionII}	
\end{figure}

This technique of `absorption' of the memory elements from a set of nodes which are the `heads' of the outgoing edges from a node $v$, to the node $v$ itself, is beneficial in terms of reducing the overall memory usage of the network (to achieve single-generation network coding) if the condition (\ref{eqn29}) is satisfied as a strict inequality.  
\begin{example}
\label{ex2}
Fig. \ref{fig:memnodereductionII} illustrates an example for memory reduction between multiple nodes ($v_1,v_2,v_3$ and $v_4$ here) of a network. Here $M_{\Gamma_{O}'(v)}=1, |\Gamma_{O}'(v)|=3$, and $M_{e_1,\Gamma_{O}'(v)}=M_{e_2,\Gamma_{O}'(v)}=1.$ Therefore, three memory elements at nodes $v_2,v_3$ and $v_4$ are `absorbed' 
into two memory elements at node $v_1.$ The boxes indicate the use of memory elements and the node to which the memory elements are attached.
\end{example}
\subsubsection{Memory reduction between nodes not necessarily adjacent}
\label{distributedmemreduction2}
For ${\cal E}_I,{\cal E}_O \subset \tilde{\cal E}$ being two sets of edges, we say that they form a pair $\left[{\cal E}_I,{\cal E}_O\right]$ if 
\begin{align*} 
{\cal E}_I=&\bigcup_{e_j\in{\cal E}_O}\Gamma_{I,e_j}(tail(e_j)).\\
&\text{and}\\
{\cal E}_O=&\bigcup_{e_i\in{\cal E}_I}\Gamma_{O,e_i}(head(e_i)).
\end{align*}
We say that the sets ${\cal E}_I,{\cal E}_O$ form a pair $\left.\left[{\cal E}_I,{\cal E}_O\right.\right)$ if 
\begin{align*} 
{\cal E}_I=&\bigcup_{e_j\in{\cal E}_O}\Gamma_{I,e_j}(tail(e_j)).\\
&\text{and}\\
{\cal E}_O\subset&\bigcup_{e_i\in{\cal E}_I}\Gamma_{O,e_i}(head(e_i)).
\end{align*}

\begin{figure}[htbp]
\centering
\includegraphics[totalheight=2.5in,width=3.4in]{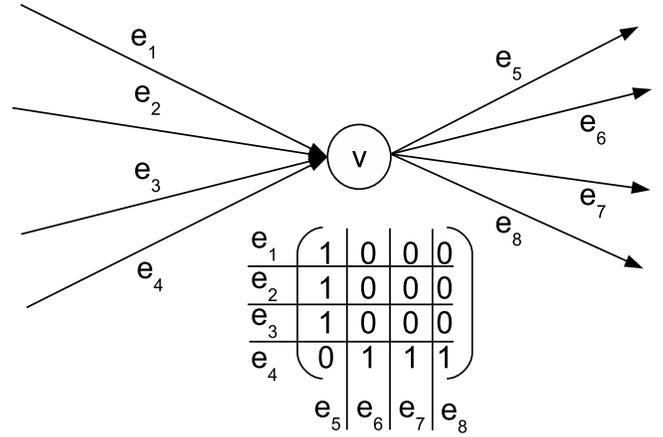}
\caption{The figure corresponding to Example \ref{example2} which gives the set $P_v$ of the node $v.$} 
\label{fig:example2}	
\end{figure}

For a node $v$, we define the set $P_v$ as follows
\[
P_v:=\left\{\left[{\Gamma}_{I_i}(v),{\Gamma}_{O_i}(v)\right]~|~1\leq i\leq s_v\right\}
\]
such that the following conditions are satisfied
\begin{align}
\label{eqn36}
&{\Gamma}_{I_i}(v)\cap{\Gamma}_{I_j}(v)&=\phi,~\forall~ 1\leq i,j \leq s_v,~i\neq j\\
\label{eqn37}
&{\Gamma}_{O_i}(v)\cap{\Gamma}_{O_j}(v)&=\phi,~\forall~ 1\leq i,j \leq s_v,~i\neq j
\end{align}
where $s_v$ is the maximum number of sets satisfying conditions (\ref{eqn36}) and (\ref{eqn37}). Algorithm \ref{alg:p_v} shown at the top of the next page obtains the set $P_v$ for some node $v.$
\begin{algorithm*}
\SetLine
\linesnumbered
\vspace{0.1cm}
\KwIn{A node $v\in\cal V$ with the edge sets $\Gamma_I(v)\cup\tilde{\Gamma}_I(v)$ and $\Gamma_O(v)\cup\tilde{\Gamma}_O(v).$}
\KwOut{The set $P_v$ for the node $v.$}
Let $i = 1, Out(v)=\Gamma_O(v)\cup\tilde{\Gamma}_O(v), P_v=\phi.$ 

\Repeat{$Out(v) = \phi$}{
Let $\Gamma_{I_i}(v)=\Gamma_{O_i}(v)=\phi.$

For some $e_j\in Out(v)$, let $\Gamma_{I_i}(v)=\Gamma_{I,e_j}(v)$

\Repeat{the sets $\Gamma_{I_i}(v)$ and $\Gamma_{O_i}(v)$ remain unchanged for $2$ consecutive iterations}
{
Let 
\[
\Gamma_{O_i}(v)=\bigcup_{e_i\in\Gamma_{I_i}(v)}\Gamma_{O,e_i}(v)
\]

Let 
\[
\Gamma_{I_i}(v)=\bigcup_{e_j\in\Gamma_{O_i}(v)}\Gamma_{I,e_j}(v)
\]
}

Let $P_v=P_v\cup\left\{\left[\Gamma_{I_i}(v),\Gamma_{O_i}(v)\right]\right\}.$

Let $Out(v)=Out(v)\backslash\Gamma_{O_i}(v)$ and $i = i + 1.$
}
\caption{Algorithm to obtain the set $P_v$ for a node $v.$}
\label{alg:p_v}
\hrule
\end{algorithm*}
\begin{example}
\label{example2}
Fig. \ref{fig:example2} illustrates a node $v$ with the local kernel matrix over some field $\mathbb{F}_q.$ For this node, the set $P_v$ is given as 
\begin{equation*}
P_v=\left\{\left[\Gamma_{I_1}(v),\Gamma_{O_1}(v)\right], \left[\Gamma_{I_2}(v),\Gamma_{O_2}(v)\right]\right\}
\end{equation*}
where 
\begin{align*}
\Gamma_{I_1}(v)=&\left\{e_1,e_2,e_3\right\}&\Gamma_{O_1}(v)=&\left\{e_5\right\}\\
\Gamma_{I_2}(v)=&\left\{e_4\right\}&\Gamma_{O_2}(v)=&\left\{e_6,e_7,e_8\right\}.
\end{align*}
\end{example}

For an pair of edge-sets $\left[{\Gamma}_{I_i}(v),{\Gamma}_{O_i}(v)\right]\in P_v$, we define $S_i(v)$, a sequence of pairs of edge-sets as
\begin{equation}
\label{eqn41}
S_i(v):=
\left.\left[{\cal E}_{i_m},{\cal E}_{i_{m-1}}\right.\right),\left[{\cal E}_{i_{m-1}},{\cal E}_{i_{m-2}}\right],...,\left[{\cal E}_{i_{2}},{\cal E}_{i_{1}}\right],\left[{\cal E}_{i_{1}},{\cal E}_{o_{1}}\right]
\end{equation}
where $\left[{\cal E}_{i_{1}},{\cal E}_{o_{1}}\right]=\left[{\Gamma}_{I_i}(v),{\Gamma}_{O_i}(v)\right],$ and $m$ is the maximum length of the sequence, that is possible to be obtained as in (\ref{eqn41}) for the edge-set pair $\left[{\Gamma}_{I_i}(v),{\Gamma}_{O_i}(v)\right].$ 

Let $k$ be an integer such that
\[
|{\cal E}_{i_{k}}|=\min_{1\leq j \leq m}|{\cal E}_{i_{j}}|.
\]

\begin{figure*}
\centering
\includegraphics[totalheight=4.5in,width=5.5in]{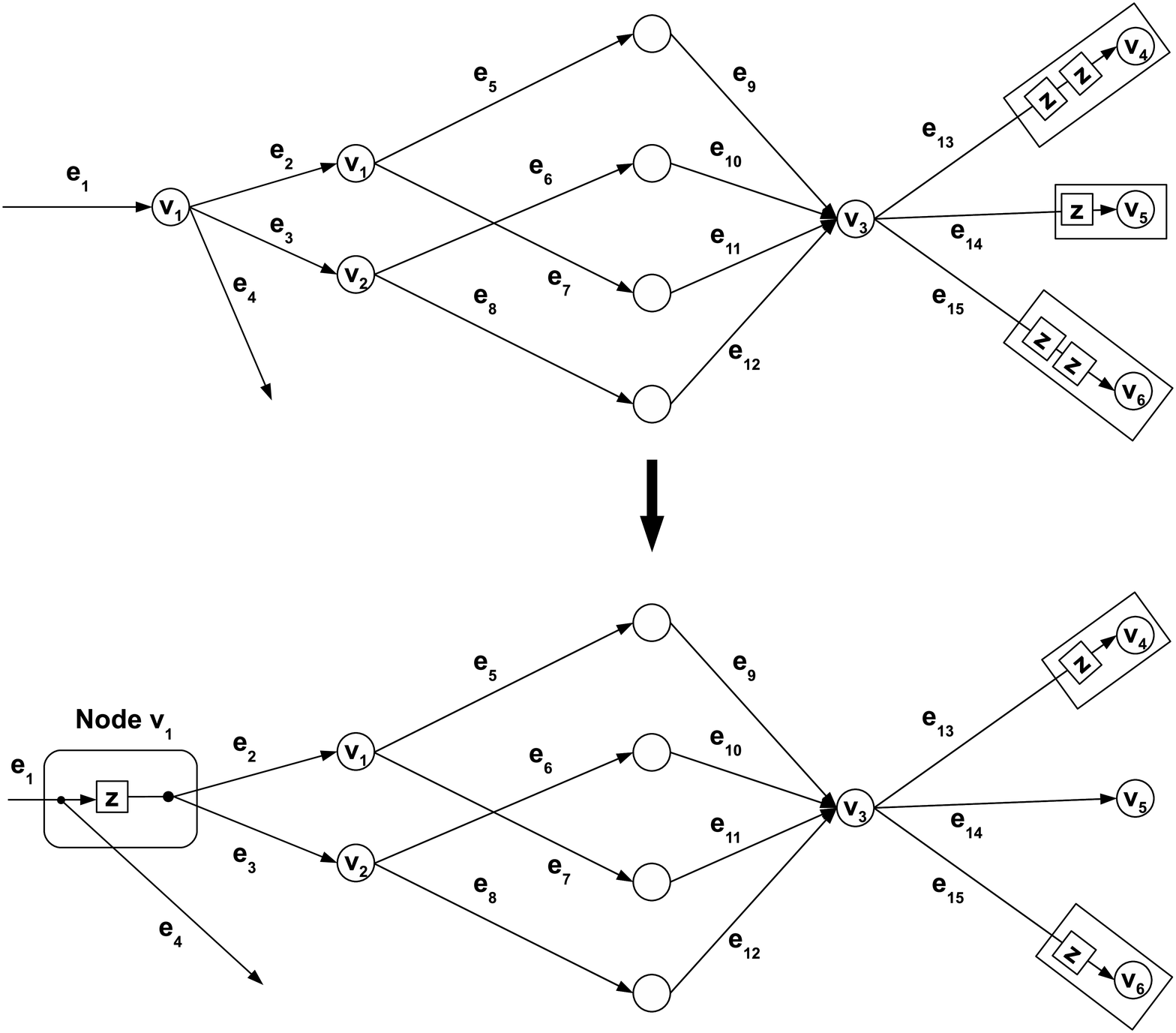}
\caption{The figure corresponding to Example \ref{example1} (Memory reduction between non-adjacent nodes).}	
\label{fig:ex5}	
\hrule
\end{figure*}

\begin{figure}[htbp]
\centering
\includegraphics[totalheight=3.5in,width=3.5in]{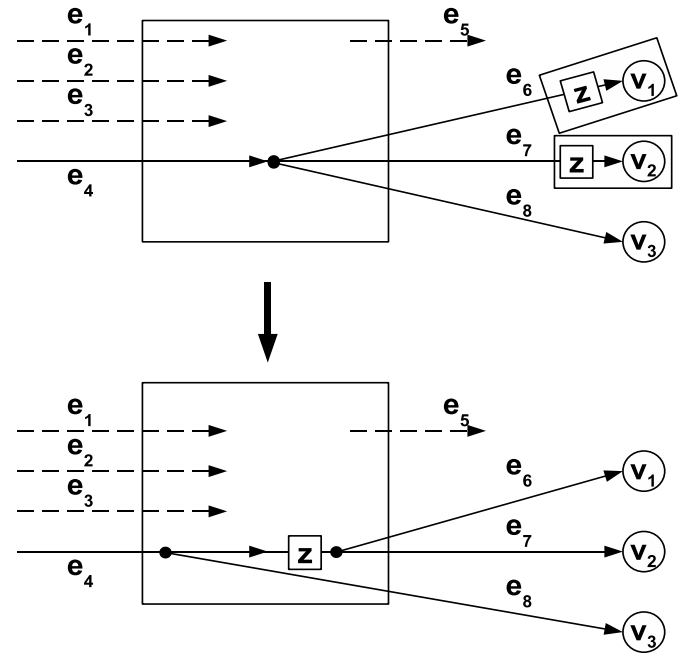}
\caption{The figure corresponding to Example \ref{example3}. The box with the incoming edges $e_1, e_2, e_3,$ and $e_4$ represents the node $v$ of Fig. \ref{fig:example2} (Example \ref{example2}).} 	
\label{fig:example3}	
\end{figure}

For the set ${\Gamma}_{O_i}(v),$ let $M_{\Gamma_{O_i}}(v)$ be defined as in (\ref{eqn39}), and the set of nodes ${\cal V}_{\Gamma_{O_i}(v)}$ be defined as in (\ref{eqn40}). Let the set of nodes ${\cal V}_{{\cal E}_{i_k}}$ be defined as in (\ref{eqn40}) for the set ${\cal E}_{i_{k}}.$ Also, let $M_{e_{i_{k}},{\Gamma_{O_i}}}(v)$ be defined as in (\ref{eqn43}) for the set ${\Gamma}_{O_i}(v)$ and for an edge $e_{i_{k}} \in {\cal E}_{i_{k}}.$ As in the memory reduction procedure of adjacent nodes, if 
\begin{equation}
\label{eqn42}
\sum_{e_{i_{k}} \in {\cal E}_{i_{k}}}M_{e_{i_{k}},\Gamma_{O_i}(v)} \leq M_{\Gamma_{O_i}}(v)|{\Gamma}_{O_i}(v)| 
\end{equation}
then the $|\Gamma_{O,i}(v)|M_{\Gamma_{O,i}(v)}$ used at the nodes ${\cal V}_{\Gamma_{O_i}(v)}$ (to delay symbols coming from the edges $e_j \in \Gamma_{O,i}(v)$) can be removed without changing the global kernels of the edges of $\Gamma_O(v'),~\forall~v' \in {\cal V}_{\Gamma_{O_i}(v)}$ by adding $M_{e_{i_{k}},\Gamma_{O_i}(v)}$ memory elements for each edge $e_{i_{k}} \in {\cal E}_{i_{k}}$ at the node $head(e_{i_{k}})\in{\cal V}_{{\cal E}_{i_k}}.$ This technique will save memory if the condition (\ref{eqn42}) is satisfied as a strict inequality.
\begin{example}
\label{example1}
Figure \ref{fig:ex5} illustrates an example for the memory reduction procedure between non-adjacent nodes. Let $K_{e_i,e_j}\neq 0,~\forall~9\leq i\leq 12,~13\leq j \leq 15.$ In the example, for the node ${v_3}$, the set $P_{v_3}$ and the sequence $S_1(v_3)$ corresponding to the only element of $P_{v_3}$ are given by (\ref{eqn47}) and (\ref{eqn48}) at the top of the next page. 
\begin{figure*}
\begin{align}
\label{eqn47}
P_{v_3}=\boldsymbol{\left\{\right.}\left[\Gamma_{I_1}(v_3)=\left\{e_{9},e_{10},e_{11},e_{12}\right\},\Gamma_{O_1}(v_3)=\left\{e_{13},e_{14},e_{15}\right\}\right]\boldsymbol{\left.\right\}}.
\end{align}
\hrule
\end{figure*}
\begin{figure*}
\begin{align}
\label{eqn48}
S_1(v_3) = \left.\left[\left\{e_1\right\},\left\{e_2,e_3\right\}\right.\right),~\left[\left\{e_2,e_3\right\},\left\{e_5,e_6,e_7,e_8\right\}\right],~\left[\left\{e_5,e_6,e_7,e_8\right\},\Gamma_{I_1}(v_3)\right],~\left[\Gamma_{I_1}(v_3),\Gamma_{O_1}(v_3)\right]
\end{align}
\hrule
\end{figure*}

Now, we have $M_{\Gamma_{O,1}(v_3)}=1$, $|\Gamma_{O,1}(v_3)|=3$, ${\cal E}_{i_{k}}=\left\{e_1\right\}$ and $M_{e_1,\Gamma_{O,1}(v_3)}=1.$ Therefore, the $3$ memory used for the edges in $\Gamma_{O,1}(v_3)$ at the nodes $v_4,v_5,$ and $v_6$ are `absorbed' into a single memory element used at node $v_1$ for edge $e_1$, thus reducing the memory usage by $2.$
\end{example}
\begin{remark}
The memory reduction procedures of Subsubsection \ref{distributedmemreduction1}, and Subsubsection \ref{distributedmemreduction2} can sometimes result in exactly the same memory reduction event. However, there could be instances in which only one of the procedures can achieve memory reduction. 

For example, the memory reduction procedure of Subsubsection \ref{distributedmemreduction1} cannot reduce memory at node $v_3$ in the situation shown in Example \ref{example1} because for any $\Gamma_{O}'(v) \subseteq \Gamma_{O}(v)$, $|\Gamma_{I}'(v)| > 3 \geq |\Gamma_{O}'(v)|$, since $\Gamma_{I}'(v)=\Gamma_{I}(v).$ However the memory reduction procedure of Subsubsection \ref{distributedmemreduction2} does work as shown in Fig \ref{fig:ex5}. 

Similarly, in some cases, at a node, the procedure of Subsubsection \ref{distributedmemreduction1} can be used to reduce memory usage, while Subsubsection \ref{distributedmemreduction2} cannot be applied. This is because of the fact that, at any node, the procedure of Subsubsection \ref{distributedmemreduction2} takes into account only those sets of the form $P_v$, while the procedure of Subsubsection \ref{distributedmemreduction1} takes into account all possible incoming and outgoing edges. Such a case is seen in Example \ref{example3}. 
\begin{example}
\label{example3}
Fig. \ref{fig:example3} shows the node $v$ of Fig. \ref{fig:example2} (Example \ref{example2}) in a particular configuration. The memory reduction procedure of Subsubsection \ref{distributedmemreduction2} cannot be applied for the set $\Gamma_{O,2}(v)$ because $M_{\Gamma_{O,2}(v)}=0.$ 

But $M_{\left\{e_6,e_7\right\}}=1$, and therefore $2$ memory elements at node $v_1$ and $v_2$ can be absorbed into a single memory element at node $v$, thereby facilitating memory reduction according to Subsubsection \ref{distributedmemreduction1}.  
\end{example}
\end{remark}
\subsection{Memory distribution}
\label{memdist}
The following technique can be used to distribute memory elements throughout the network in a somewhat uniform way. Suppose there exists a node $v\in \cal V$ such that for some $e_j \in \Gamma_O(v)$ with $v' = head(e_j)$ and for some integer $m\leq M_{e_j,head(e_j),min}$,
\begin{equation}
\label{eqn30}
M_v+m\leq M_{v'}-m
\end{equation}
then the $m$ memory elements at node $v'$ used to delay symbols coming from edge $e_j$ can be `absorbed' into node $v$ (thereby using them to delay symbols going into edge $e_j$) without changing the global kernels of any edge in $\Gamma_O(v').$ 

This technique reduces the number of memory elements used at node $v'$ for delaying its incoming symbols while increasing the number ($M_{e_j,tail(e_j)}$) of memory elements used at node $v$ for delaying its outgoing symbols. 

\begin{figure}[htbp]
\centering
\includegraphics[totalheight=2.9in,width=3.2in]{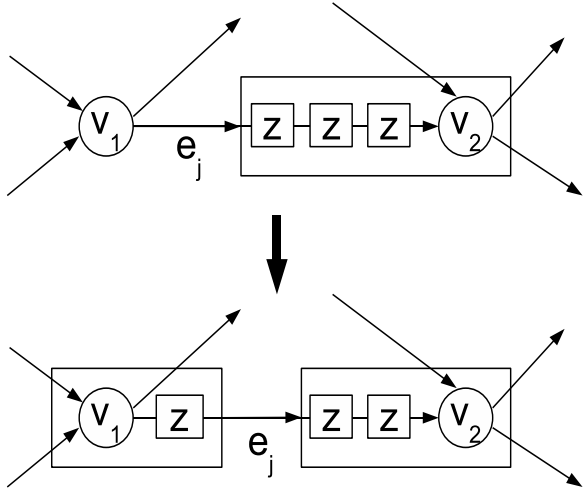}
\caption{The figure corresponding to Example \ref{ex3} (Memory distribution).} 
\label{fig:memdist}	
\end{figure}

\begin{example}
\label{ex3}
Fig \ref{fig:memdist} illustrates an example for memory distribution between two nodes $v_1$ and $v_2.$ In the figure on the top, $m=1, M_{v_1}=0,$ and $M_{v_2}=3.$ Therefore one memory element from $v_2$ (used to delay symbols coming from $e_j$ into $v_2$) can be `absorbed' into node $v_1$ (and thereby used to delay symbols going into $e_j$ from $v_1$). The boxes indicate the node to which the memory elements are attached. After distribution, $M_{v_1}=1,$ and $M_{v_2}=2.$
\end{example}
\section{Single-generation network coding - Algorithm}
\label{sec5}

This section presents the main contribution of this paper. 

For an edge $e_i \in {\cal E}$, let $\boldsymbol{f}_{e_i}(z)\in\mathbb{F}_q^n(z)$ represent the global kernel of $e_i.$ We say that a node $v \in {\cal V}\backslash\left\{s\right\}$ is a \textit{coding node} if the global kernel of at least one of its outgoing edge is a $\mathbb{F}_q(z)$ linear combination of the global kernels of at least two of its incoming edges. Otherwise, we call $v$ a \textit{forwarding node}.

Let ${\cal V}_{cod}$ be the set of coding nodes, and ${\cal V}_{fwd}$ be the set of forwarding nodes. Let ${\cal V}_{cod}^{0}$ be the set of all coding nodes such that there exist no path in the network from any other coding node to any node in ${\cal V}_{cod}^{0}.$ 

Towards proposing an algorithm to enable single-generation network coding, we make some observations and discuss the addition of memory elements at the coding nodes to achieve single-generation network coding. 
\begin{observation}
For any $v \in {\cal V}_{cod}^{0}$, the global kernel of any $e \in \Gamma_I(v)$ is of the form
\begin{equation}
\label{eqn23}
\boldsymbol{f}_e(z)=z^{l_e}\boldsymbol{f}_e
\end{equation}
for some positive integer $l_e$, with $\boldsymbol{f}_e \in \mathbb{F}_q^n.$ If the network is a unit-delay network and the node $v$ uses no memory, the global kernel of any $e_j \in \Gamma_O(v)$ is of the form
\begin{equation}
\label{eqn24}
\boldsymbol{f}_{e_j}(z)=\sum_{e_i \in \Gamma_I(v)}zK_{e_i,e_j}\boldsymbol{f}_{e_i}(z)=\sum_{e_i \in \Gamma_I(v)}K_{e_i,e_j}z^{l_{e_i}+1}\boldsymbol{f}_{e_i}
\end{equation}
where $l_{e_i}$ is a positive integer signifying accumulated delay from the source to edge $e_i$, and $K_{e_i,e_j} \in \mathbb{F}_q$ signifies the local kernel coefficient between $e_i$ and $e_j.$ The additional $z$ is to account for the delay in the unit delay network.
\end{observation}
\subsection{Single-generation processing at the nodes}
\label{singlegennode}
For every pair of edges $e_i, e_{i'}\in \Gamma_{I,e_j}(v)$ ($\Gamma_{I,e_j}(v)$ being as in (\ref{eqn27})) in (\ref{eqn24}) such that $l_{e_i} < l_{e_{i'}}$, we may add  $M_{e_i,e_j}=l_{e_{i'}}-l_{e_i}$ memory elements at node $v$ to delay the symbols coming from $e_i$ such that the global kernel of the edge $e_j$ becomes
\begin{equation}
\label{eqn44}
\boldsymbol{f}_{e_j}(z)=z^{l_{e_j,max}+1}\sum_{e_i \in \Gamma_I(v)}K_{e_i,e_j}\boldsymbol{f}_{e_i}
\end{equation}
where $l_{e_j,max} = \max_{e_i \in \Gamma_{I,e_j}(v)}l_{e_i}$ and $K_{e_i,e_j} \in \mathbb{F}_q.$ Once this process of using memory at the node $v$ results in the global kernel of every edge in $\Gamma_O(v)$ to be a linear combination of symbols from the same generation (generations between different outgoing edges need not be the same), we say that \textit{single-generation processing} has been achieved at node $v.$ For a node $T \in \cal T$, we	say that single-generation processing has been achieved at sink $T$ if the condition (\ref{eqn28}) is satisfied along with condition (\ref{eqn44}) for each $e_j\in\Gamma_O(T).$
\begin{observation}
We iteratively define the set ${\cal V}_{cod}^{i} \subseteq {\cal V}_{cod}$ as the set of coding nodes which have path only from  
\[
\left(\bigcup_{j=0}^{i-1}{\cal V}_{cod}^{j}\right)\bigcup{\cal V}_{fwd}
\]
where ${\cal V}_{cod}^{0}$ is as defined before. Once memory has been used to achieve single-generation processing at all nodes in ${\cal V}_{cod}^{i-1}$, it can be observed that the global kernels of the incoming and outgoing edges of any node $v \in {\cal V}_{cod}^{i}$ satisfy the same condition as in (\ref{eqn23}) and (\ref{eqn24}). 
\end{observation}

Thus again memory elements can be used at the nodes of ${\cal V}_{cod}^{i}$ to implement single-generation processing, ultimately achieving single-generation processing at each coding node of the network. 
\subsection{Algorithm for single-generation network coding}
Algorithm \ref{alg:singlegen} shown in the next page is used to achieve single-generation network coding using memory at the nodes of the network, while trying to minimize the total number of memory elements used in the network.
\begin{algorithm*}
\SetLine
\linesnumbered
\vspace{0.1cm}
\KwIn{A network ${\cal G}_m$ with delays and unused memory elements}
\KwOut{The network ${\cal G}_m$ with a single-generation network code using memory elements at nodes}
\ForEach{$v \in {\cal V}_{cod}$ in the ancestral order}
{
Introduce sufficient memory elements at node $v$ accordingly as in Subsection \ref{singlegennode} in order to enable single-generation processing at node $v.$

\ForEach{$e_i \in \Gamma_I(v)\cup\tilde{\Gamma}_I(v)$}
{
Run the memory reduction procedure as in Subsection \ref{nodememreduction}.
}
}

Now the global kernel of any edge $e_j \in \Gamma_I(T)$ of any sink $T$ is of the form
\[
\boldsymbol{f}_{e_j}(z)= z^{L_{e_j}}\boldsymbol{f}_{e_j}
\]
for some positive integer $L_{e_j},$ with $\boldsymbol{f}_{e_j} \in \mathbb{F}_q^n.$

\ForEach{$T\in\cal T$}
{ 
Add sufficient memory according to Subsection \ref{pairmem} and Subsection \ref{outmem} such that single-generation processing is achieved at the sink $T.$
}

\ForEach{$v \in {\cal V}$ in the reverse-ancestral order}
{
	\ForEach{pair of edge-sets $\left[\Gamma_{I_i}(v), \Gamma_{O_i}(v)\right] \in P_v$}{
		\If{condition (\ref{eqn42}) is satisfied}
		{
		Run the memory reduction procedure as in Subsubsection \ref{distributedmemreduction2}.
		}
	}
}

\ForEach{$v \in {\cal V}$ in the reverse-ancestral order}
{
	\ForEach{subset $\Gamma_{O}'(v) \subseteq  \Gamma_{O}(v)\cup\tilde{\Gamma}_{O}(v)$}{
		\If{condition (\ref{eqn29}) is satisfied}
		{
		Run the memory reduction procedure as in Subsubsection \ref{distributedmemreduction1}.
		}
	}
}

\ForEach{$v \in {\cal V}$ in the ancestral order}
{
	\ForEach{$e_j \in \Gamma_O(v)$}
	{
		\If{condition (\ref{eqn30}) is satisfied}
		{
		Run the memory distribution procedure at $v$ as in Subsection \ref{memdist}.
		}
	}
}

\ForEach{$v \in {\cal V}$ in the ancestral order}
{
	\ForEach{$e_j \in \Gamma_O(v)\cup\tilde{\Gamma}_O(v)$}
	{
		\ForEach{$e_i \in \Gamma_{I,e_j}(v)$}
		{
			Update the corresponding elements in $A$, $K$, and $B^v$ matrices according to (\ref{eqn31}), (\ref{eqn21}), and (\ref{eqn32}) of Subsection \ref{pairmem} upon calculating $M_{e_i,e_j}.$
		}
		Update the corresponding elements in $A$, $K$, and $B^v$ matrices according to (\ref{eqn33}), (\ref{eqn22}), and (\ref{eqn34}) of Subsection \ref{outmem} upon calculating $M_{e_j,tail(e_j)}.$
	}
}
\caption{Algorithm for using memory at nodes to obtain a single-generation network code}
\label{alg:singlegen}
\hrule
\end{algorithm*}
\begin{remark}
\label{remark1}
Algorithm \ref{alg:singlegen} assumes that every node has unlimited memory to use and then tries to obtain a configuration that reduces the number of memory elements used in the network. However, if the maximum available memory in the nodes is limited, then the following techniques may be adopted after running Algorithm \ref{alg:singlegen}.
\begin{itemize}
\item In line 27 of the algorithm, instead of checking condition (\ref{eqn30}) at every pair of nodes connected by some edge, the actual memory capability of the nodes must be taken into account and then the distribution procedure of Subsection \ref{memdist} can be run.
\item Finally, at every node in which the algorithm demands more memory elements than what is available, sufficient memory elements should be removed so that the total memory used at the node is utmost what is available. As the penalty of removing these memory elements will be compensated by the sinks, the memory elements that will be removed at the nodes should ideally be such that the compensation occurs in the least number of sinks in the least possible quantity.
\end{itemize}
\end{remark}
\begin{example}
\label{lastex}
Fig. \ref{fig:ex2a}, Fig. \ref{fig:ex3}, and Fig. \ref{fig:ex4} represent the network at various stages of the algorithm applied on a modified double-butterfly network as shown in Fig. \ref{fig:ex1a}. The modified unit-delay double-butterfly network shown in Fig. \ref{fig:ex2a} has the standard network code over $\mathbb{F}_2.$ $s$ is the source node, $T_i, i=1,2,3,4$ are the sinks. The dotted lines represent the virtual input edges at the source and virtual output edges at the sinks.

\begin{figure*}
\centering
\includegraphics[totalheight=3in,width=6in]{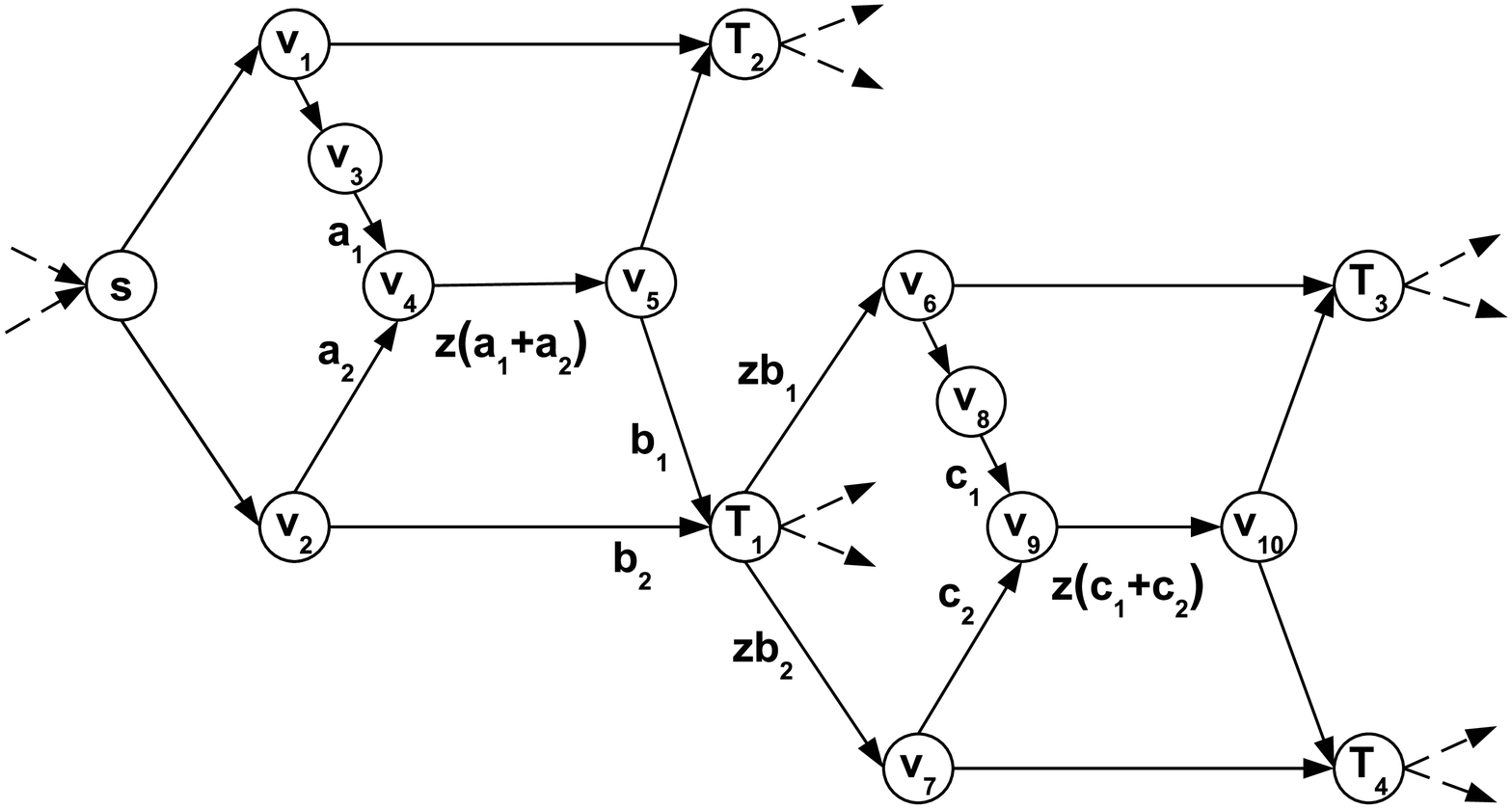}
\caption{Figure corresponding to Example \ref{lastex}. A modified double-butterfly network. The mapping between the incoming and outgoing symbols ($a_1,a_2,b_1,b_2,c_1,c_2 \in \mathbb{F}_2$) at the nodes $v_4$, $T_1$, and $v_9$ are shown.}
\label{fig:ex1a}	
\hrule
\end{figure*}
%
\begin{figure*}
\centering
\includegraphics[totalheight=5in,width=6in]{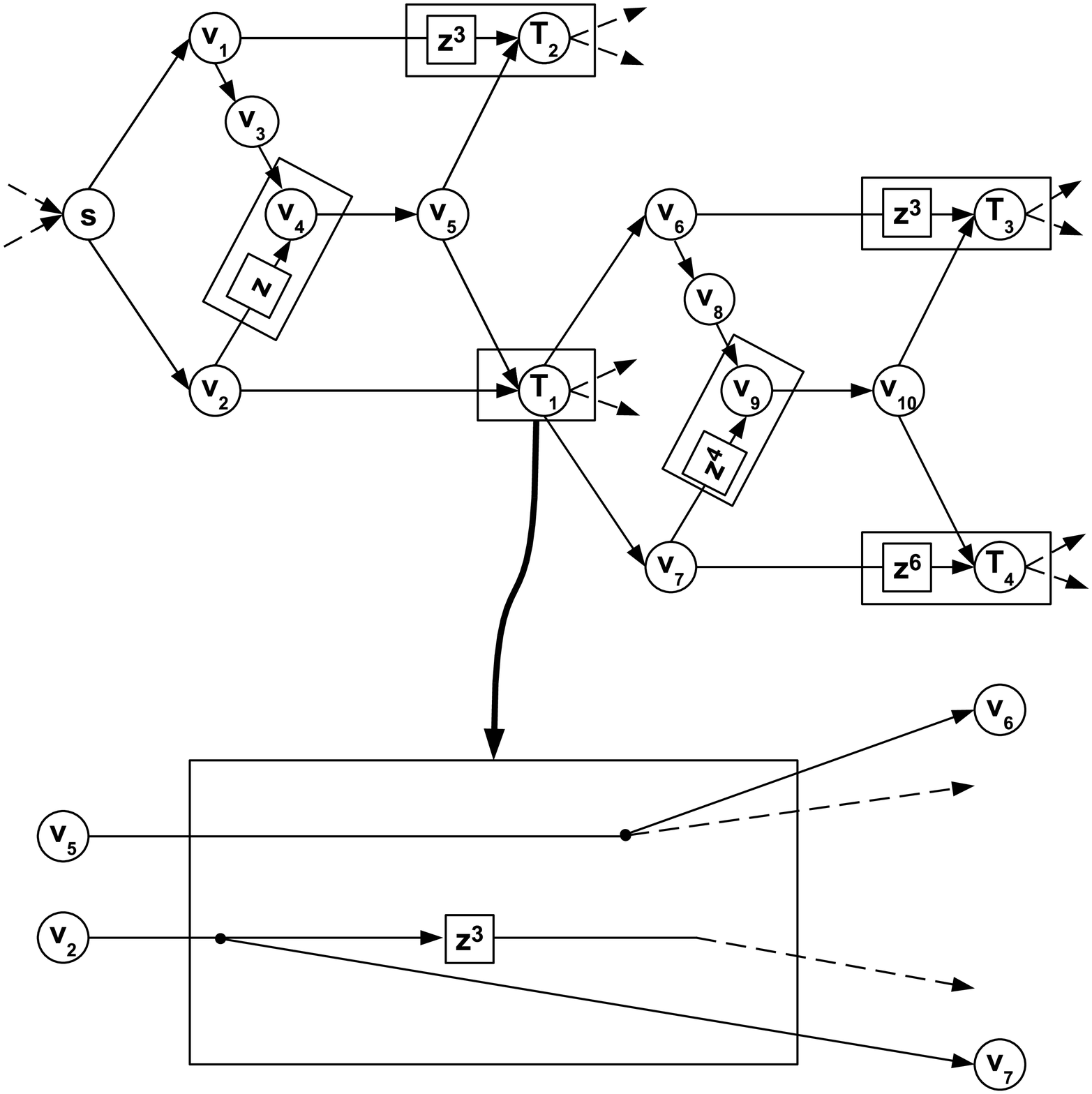}
\caption{Figure corresponding to Example \ref{lastex}. After line 10 of Algorithm \ref{alg:singlegen}, single-generation network coding has been implemented in the network and all the sinks see a network transfer matrix as in (\ref{eqn28}). Each box indicates the presence of memory elements at the associated node. The way sink $T_1$ uses memory is expanded below. Total memory used at this stage is 20.}	
\label{fig:ex2a}	
\hrule
\end{figure*}
\begin{figure*}
\centering
\includegraphics[totalheight=3in,width=6in]{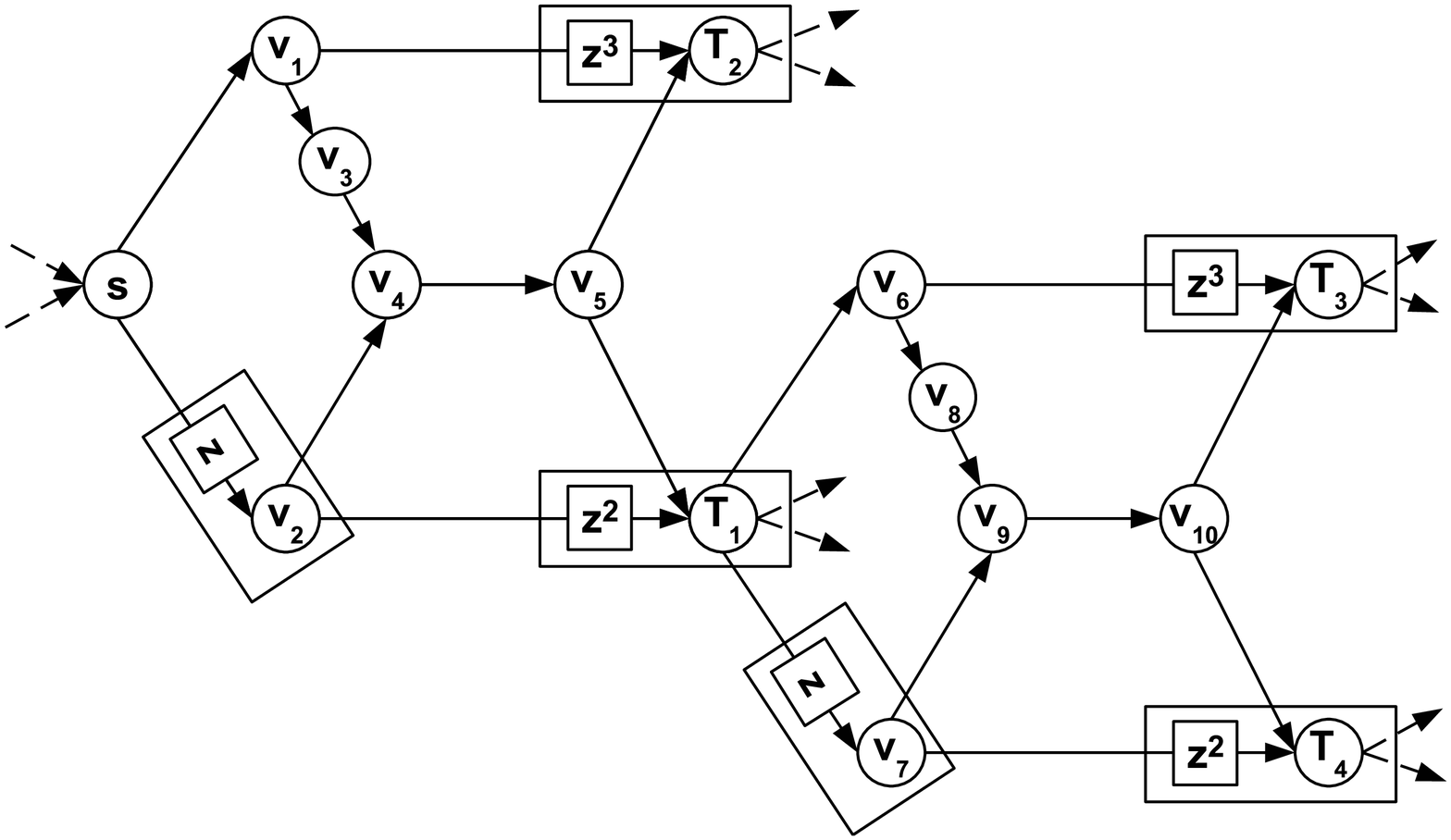}
\caption{Figure corresponding to Example \ref{lastex}. The network after line 24 of the algorithm. Comparing this figure with Fig. \ref{fig:ex2a}, memory reduction according to Subsubsection \ref{distributedmemreduction1} has resulted in the `absorption' of memory elements from the nodes $v_4, T_1, v_7, v_9,$ and $T_4.$ Total memory used in the network now is 12.}	
\label{fig:ex3}	
\hrule
\end{figure*}
\begin{figure*}
\centering
\includegraphics[totalheight=3in,width=6in]{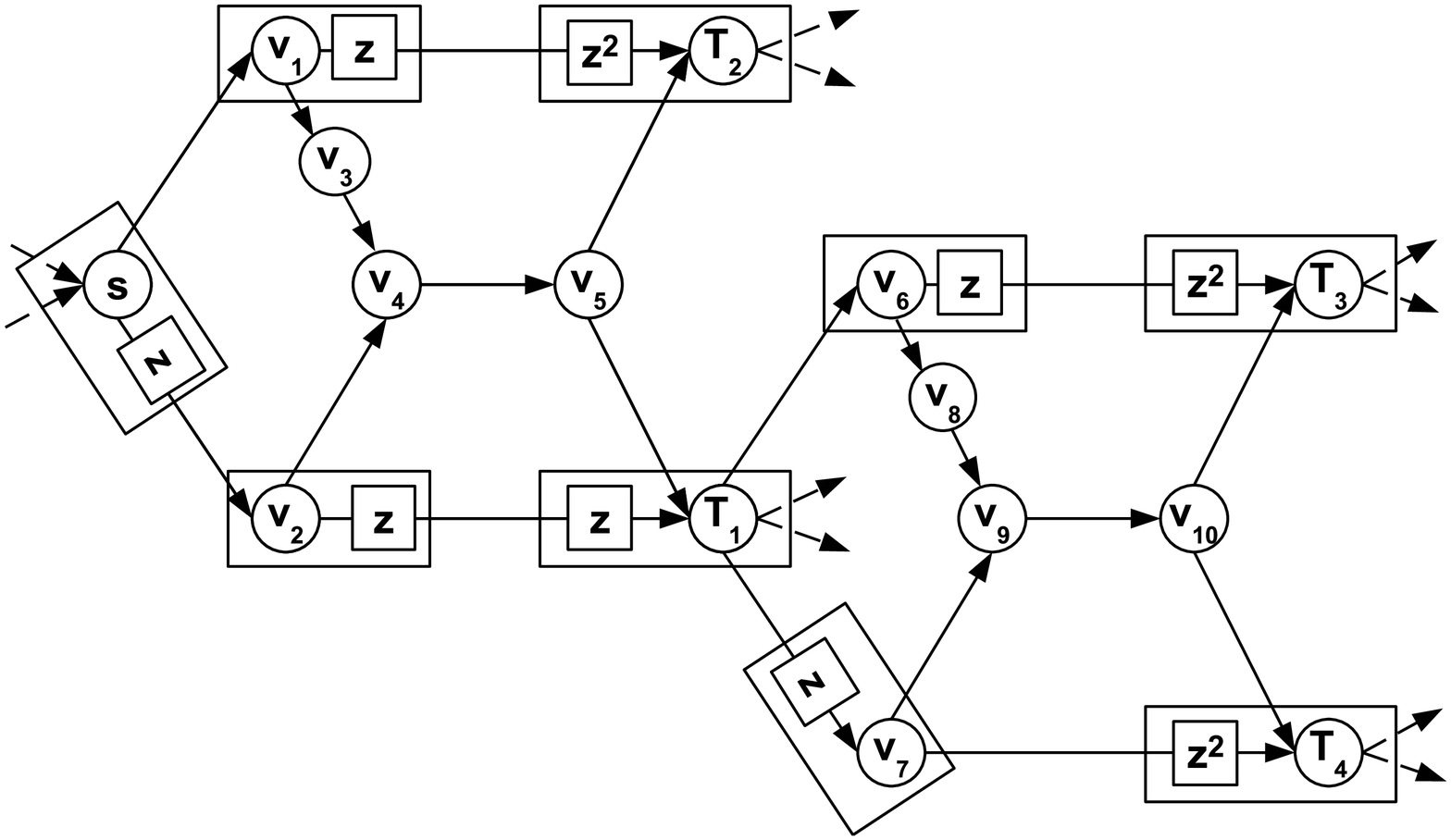}
\caption{Figure corresponding to Example \ref{lastex}. The network at the end of Algorithm \ref{alg:singlegen}. The $12$ memory elements used in Fig. \ref{fig:ex3} are further distributed amongst the nodes of the network.}	
\label{fig:ex4}	
\hrule
\end{figure*}

Table \ref{tab1} shows the network transfer matrices before and after obtaining single-generation processing using Algorithm \ref{alg:singlegen}. Table \ref{tab1} also shows a comparison between the memory requirements at the sinks (for decoding) between inter-generation network coding (i.e the memory-free case; the numbers shown are the sum of the row degrees of realizable inverse matrices in the third column) and single-generation network coding (as shown in Fig. \ref{fig:ex4}). In the memory-free case, assuming that sinks use memory individually to decode, the total number of memory elements used in the network is $19$, and all of them are used at the sinks. In the single-generation network coded network as shown in Fig. \ref{fig:ex4}, it can be seen that the total number of memory elements used in the network is $12$, out of which only $7$ are used at the sinks, thereby showing a marked reduction from the memory-free case. The rest of the memory elements (numbering $5$) are distributed across the nodes of the network. 
\end{example}
\begin{table*}
\centering
\caption{Comparing inter(memory-free) and single-generation network coding(using memory) for the network in Fig. \ref{fig:ex1a}}
\begin{tabular}{|c|c|c|c|c|c|}\hline
\textbf{Sink} & \textbf{Network transfer matrix} & \textbf{Realizable decoding matrix} & \textbf{Network transfer matrix} & \textbf{No. of memory} & \textbf{No. of memory}\\
& \textbf{before Algorithm \ref{alg:singlegen}} &  \textbf{obtained from $M_T^{-1}(z)$} & \textbf{after Algorithm \ref{alg:singlegen}} &  \textbf{elements used} &  \textbf{elements used}\\
&&&& \textbf{before Algorithm \ref{alg:singlegen}} & \textbf{after Algorithm \ref{alg:singlegen}}\\
\hline
$T_1$ & $M_{T_1}(z)=\left( \begin{array}{cc}z & z^3 \\0  & z^4 \end{array} \right)$ & $P_{T_1}(z)= \left( \begin{array}{cc}z^3 & z^2 \\0  & 1 \end{array} \right)$ & $M_{T_1}(z)=z^4\left( \begin{array}{cc}1 & 1 \\0  & 1 \end{array}\right)$ & 3 & 1 \\
\hline
$T_2$ & $M_{T_2}(z)=\left( \begin{array}{cc}z^3 & 0 \\z^4  & z \end{array} \right)$ & $P_{T_2}(z)= \left( \begin{array}{cc}1 & 0 \\z^3  & z^2 \end{array} \right)$ & $M_{T_2}(z)=z^4\left( \begin{array}{cc}1 & 0\\1  & 1 \end{array}\right)$ & 3 & 2 \\
\hline
$T_3$ & $M_{T_3}(z)=\left( \begin{array}{cc}z^5+z^8 & z^5  \\z^9 & z^6 \end{array} \right)$ & $P_{T_3}(z)= \left( \begin{array}{cc}z & z^4 \\1 & 1+z^3 \end{array} \right)$ & $M_{T_3}(z)=z^9\left( \begin{array}{cc}1 & 0 \\1 & 1\end{array}\right)$ & 7 & 2 \\
\hline
$T_4$ & $M_{T_4}(z)=\left( \begin{array}{cc}z^3 & z^5+z^8 \\0  & z^9 \end{array} \right)$ & $P_{T_4}(z)= \left( \begin{array}{cc}z^6 & z^2+z^5 \\0  & 1 \end{array} \right)$ & $M_{T_4}(z)=z^9\left( \begin{array}{cc}1 & 1 \\ 0  & 1 \end{array}\right)$ & 6 & 2 \\
\hline
\end{tabular}
\label{tab1}
\end{table*}
\subsection{Comparison with the approach of \cite{WZY}}
We can compare the straightforward approach of \cite{WZY} and our approach to obtaining a single-generation network coded network for the modified unit-delay double-butterfly network of Fig. \ref{fig:ex1a}. According to the technique in \cite{WZY}, the result would be the network as in Fig. \ref{fig:ex2a}, thereby resulting in the use of $20$ memory elements to obtain single-generation network coding. However, our algorithm utilizes the memory reduction and distribution techniques as given in Section \ref{sec4} and results in the output being as in Fig \ref{fig:ex4} using $12$ memory elements and a more uniform distribution of memory elements across the network than in Fig. \ref{fig:ex2a}. Although the overall memory usage is reduced, it still remains to be shown whether Algorithm \ref{alg:singlegen} actually obtains a configuration of the network with minimal number of memory elements being used to obtain single-generation network coding.
\section{Impact of single-generation network coding on network-error correction}
\label{sec6}
\subsection{Impact on encoding}
\textit{Construction of a CNECC:} For details on the basics of convolutional codes, we refer the reader to \cite{JoZ}. The construction of a CNECC \cite{PrR2} for a given acyclic, unit-delay, memory-free network which corrects error vectors corresponding to a given set $\Phi$ of error patterns (an error pattern is a subset of $\cal E$ indicating the edges in error) can be summarized as follows
\begin{itemize}
\item Compute the set ${\cal W}_s$ of \textit{error vector reflections} given by
\[
{\cal W}_s = \bigcup_{T\in{\cal T},\rho \in \Phi} \left\{\boldsymbol{w}F_T(z)p_{_T}(z)M_T^{-1}(z)~|~\boldsymbol{w}\in \rho\right\}
\]
where $\boldsymbol{w}\in F_q^{|{\cal E}|}$ is an error vector, and $\boldsymbol{w}\in \rho$ means that $\boldsymbol{w}$ matches an error pattern $\rho.$ $p_{_T}(z) \in \mathbb{F}_q[z]$(the ring of polynomials) is some \textit{processing function} chosen such that the \textit{processing matrix} $p_{_T}(z)M_T^{-1}(z)=P_T(z)$ is a polynomial matrix.
\item Let $t_s = \max_{\boldsymbol{w}_s(z) \in {\cal W}_s}w_H\left(\boldsymbol{w}_s(z)\right).$ Choose an input convolutional code ${\cal C}_s$ with free distance at least $2t_s+1$ as the CNECC for the given network. 
\end{itemize}

The following lemma gives a bound on $t_s$ and therefore the free distance demanded of the CNECC.
\begin{lemma}[ \cite{PrR2} ]
\label{tdelay}
Given an acyclic, unit-delay, memory-free network ${\cal G}({\cal V},{\cal E})$ with a given error pattern set $\Phi$, let $T_{delay}-1	$ be the maximum degree of any polynomial in the $F(z)$ matrix. Let $w_H$ indicate the Hamming weight over $\mathbb{F}_q.$ If $r$ is the maximum number of non-zero coefficients of the polynomials $p_{_T}(z)$ corresponding to all sinks in $\cal T$, i.e $r=\max_{T \in {\cal T}} w_H\left(p_{_T}(z)\right)$, then we have 
\[
t_s \leq rn\left[\left(n+1\right)\left(T_{delay}-1\right)+1\right].
\]
\end{lemma}

Algorithm \ref{alg:singlegen} does not increase the value of $T_{delay}$ in the matrix $F(z)$ because of the fact that an additional delay would not be introduced on any path between nodes which are at a distance of $T_{delay}$ edges (the maximum number of edges on any path between any two nodes) from each other. Also, with memory being introduced in the nodes according to Algorithm \ref{alg:singlegen}, the network transfer matrices at all the sinks are of the form as given in (\ref{eqn28}). Therefore the processing functions at any sink $T$ is of the form $p_{_T}(z)=z^{L_T}$, i.e $r=1.$ 

Therefore we have that, for the network with delay and memory (used to achieve single-generation network coding), 
\[
t_s \leq n\left[\left(n+1\right)\left(T_{delay}-1\right)+1\right].
\]

Thus, it is seen that the bound for $t_s$ and therefore for the free distance demanded of the CNECC may be lower (if $r > 1$) for the unit-delay, single-generation network coded network compared to the unit-delay, memory-free counterpart. However a decrease in the actual value of $t_s$ cannot be guaranteed and has to be computed for every network individually in order to decide whether the CNECC designed for the unit-delay, memory-free network will continue to work for the single-generation network coded unit-delay counterpart. 
\subsection{Impact on decoding}

\textit{Decoding of a CNECC:} Let $G_I(z)$ be the generator matrix of the code ${\cal C}_s$ thus designed. Then we refer to the code ${\cal C}_s$ as the \textit{input convolutional code} \cite{PrR}. The effective code seen by a sink $T$ is generated by the matrix $G_{O,T}(z)=G_T(z)M_T(z)$, which is known as the \textit{output convolutional code}\cite{PrR}, ${\cal C}_{O,T}$, at sink $T.$ The decoding of the CNECC at any sink $T$ can be performed either on the trellis of the code ${\cal C}_s$ or that of the code ${\cal C}_{O,T}$ at that particular sink according to the free distance of ${\cal C}_{O,T}$ ($d_{free}({\cal C}_{O,T})$), the catastrophic/non-catastrophic nature of $G_{O,T}(z)$, and a parameter called $T_{d_{free}}({\cal C}_{O,T})$, whose definition for a rate $b/c$ code $\cal C$ over $\mathbb{F}_q$ is given in \cite{PrR} as follows.
\begin{equation}
\label{tdfreedef}
T_{d_{free}}({\cal C}):=\max_{\boldsymbol{v}_{[0,j)} \in S_{d_{free}}}j+1
\end{equation}
where $S_{d_{free}}$ \cite{PrR} is defined as follows. 
\begin{equation*}
\label{sdfree}
S_{d_{free}}:=\left\{\boldsymbol{v}_{[0,j)} \mid w_H\left(\boldsymbol{v}_{[0,j)}\right) < d_{free}({\cal C}),  \boldsymbol{\sigma}_0=\boldsymbol{0},\forall~j>0 \right\}
\end{equation*}
where 
\[
\boldsymbol{v}_{[0,j)}:=\left[\boldsymbol{v}_0,\boldsymbol{v}_1,. . . ,\boldsymbol{v}_{j-1}\right]
\]
is a truncated codeword sequence with $\boldsymbol{v}_i \in \mathbb{F}_q^c)$, $\boldsymbol{\sigma}_t$ indicates the content of the delay elements in the encoder at a time $t$, and $w_H$ indicates the Hamming weight over $\mathbb{F}_q.$ The set $S_{d_{free}}$  consisting of all possible truncated codeword sequences $\boldsymbol{v}_{[0,j)}$ of weight less than $d_{free}({\cal C})$ that start in the zero state. 
%
%
Then, we have the following proposition.
\begin{proposition}[\cite{PrR} ]
\label{minweighttime}
The minimum Hamming weight trellis decoding algorithm can correct all error sequences which have the property that the Hamming weight of the error sequence in any consecutive $T_{d_{free}}({\cal C})$ segments (a segment being a collection of $c$ output symbols corresponding to every $b$ input symbols) is utmost $\left\lfloor \frac{d_{free}({\cal C})-1}{2} \right\rfloor$.
\end{proposition}

With the CNECC in place in a unit-delay. memory-free network, under certain conditions (see Subsection IV-D of \cite{PrR2}), a sink has to decode on the trellis of the input convolutional code, in which case the sink has to multiply the incoming $n$ output streams with the processing matrix $P_T(z)$, which may require additional memory elements to implement. However, with a single-generation network code implemented using memory elements, part of this processing is done in a distributed manner in the other nodes of the network, thereby decreasing the memory requirement at the sinks. 

In the forthcoming section, we further observe the advantages that the use of memory in the intermediate nodes offers in the performance of CNECCs under a probabilistic error setting.
\section{Simulation results}
\label{sec7}
\subsection{A probabilistic error model}
Probabilistic error models have been considered in the context of random network coding in \cite{SKK2}. We define a probabilistic error model for a unit delay network  ${\cal G}({\cal V},{\cal E})$ by defining the probabilities of any set of $i~(i\leq|{\cal E}|)$ edges of the network being in error at any given time instant. Across time instants, we assume that the network errors are i.i.d. according to this distribution.
\begin{align}
\label{eq:1}
Prob.&(i~\text{network edges being in error}) = p^i \\
\label{eq:2}
Prob.&(\text{no edges are in error}) = q
\end{align}
where $1 < i \leq |{\cal E}|,$ and $p,q \leq 1$ are real numbers indicating the probability of any single edge error in the network and the probability of no edges in error respectively, such that $q + \sum_{i=1}^{|{\cal E}|}p^i = 1.$
\subsection{Simulations on the modified butterfly network}
\begin{figure*}
\centering
\includegraphics[totalheight=2.5in,width=6in]{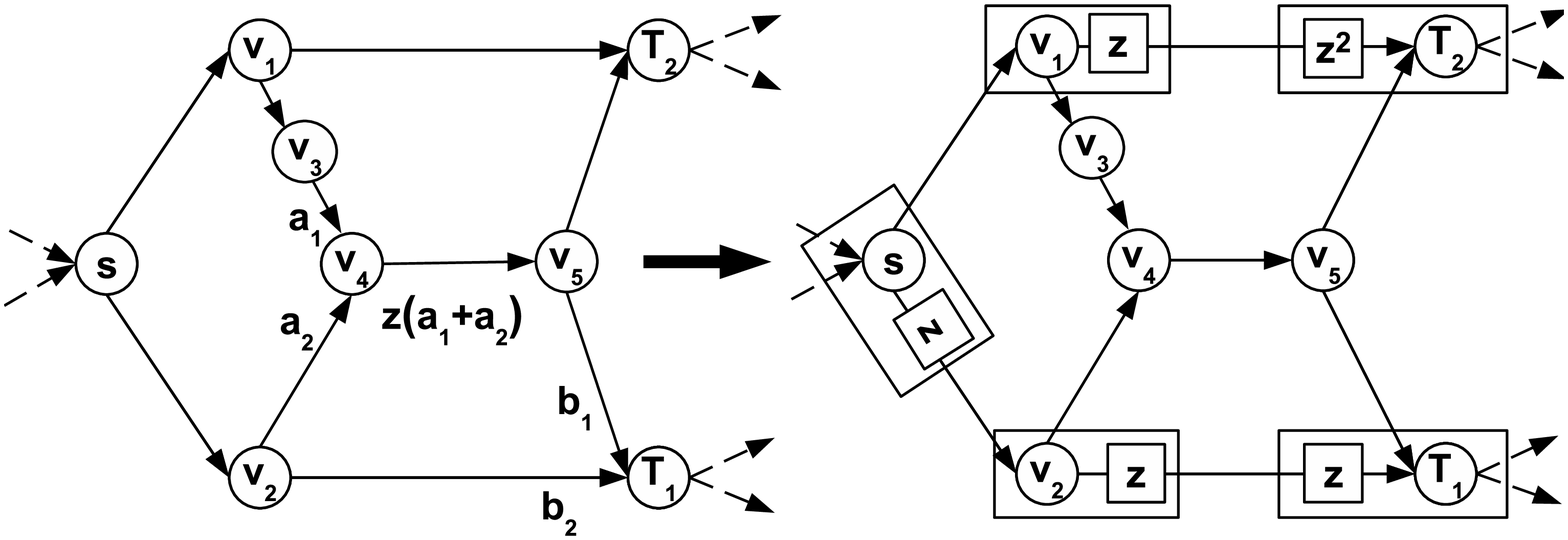}
\caption{A modified butterfly network}	
\label{fig:butterflydelaymem}	
\hrule
\end{figure*}

Fig. \ref{fig:butterflydelaymem} on the top of the next page shows a modified butterfly network before and after running Algorithm \ref{alg:singlegen}. This network is clearly a part of the modified double-butterfly network of Fig. \ref{fig:ex1a}, and the associated matrices at the sinks $T_1$ and $T_2$  are given in Table \ref{tab1}.
With the probability model as in  (\ref{eq:1}) and (\ref{eq:2}) with $|{\cal E}|=10$ for this network, we simulate the performance of $3$ input convolutional codes implemented on this network  for both the with-memory and memory-free cases as in Fig. \ref{fig:butterflydelaymem} with the sinks performing hard decision decoding on the trellis of the input convolutional code. 

In the following discussion we refer to sinks $T_1$ and $T_2$ of Fig. \ref{fig:butterflydelaymem} as Sink 1 and Sink 2. The $3$ input convolutional codes and the rationality behind choosing them are given as follows. 
\begin{itemize}
\item Code ${\cal C}_1$ is generated by the generator matrix 
\[
G_{I_1}(z)=\left[1+z~~~1\right],
\]
with $d_{free}({\cal C}_1)=3$ and $T_{d_{free}}({\cal C}_1)=2.$ This code is chosen only to illustrate the error correcting capability of codes with low values of $d_{free}({\cal C})$ and $T_{d_{free}}({\cal C}).$ 
\item Code ${\cal C}_2$ is generated by the generator matrix 
\[
G_{I_2}(z)=\left[1+z^2~~~1+z+z^2\right],
\]
with $d_{free}({\cal C}_2)=5$ and $T_{d_{free}}({\cal C}_2)=6.$ This code corrects all double edge errors in the instantaneous version (with all edge delays and memories being zero) of Fig. \ref{fig:butterflydelaymem} as long as they are separated by $6$ network uses.
\item  Code ${\cal C}_3$ is generated by the generator matrix 
\[
G_{I_3}(z)=\left[1+z+z^4~~~1+z^2+z^3+z^4\right],
\]
with $d_{free}({\cal C}_3)=7$ and $T_{d_{free}}({\cal C}_3)=12.$ This code corrects all double edge errors in the unit-delay network given in Fig. \ref{fig:butterflydelaymem} as long as they are separated by $12$ network uses.
\end{itemize}
We note here that values of $T_{d_{free}}({\cal C})$ of the $3$ codes are directly proportional to their free distances, i.e, the code with greater free distance has higher $T_{d_{free}}({\cal C})$. 

\begin{figure*}
\centering
\includegraphics[totalheight=4.9in,width=7.1in]{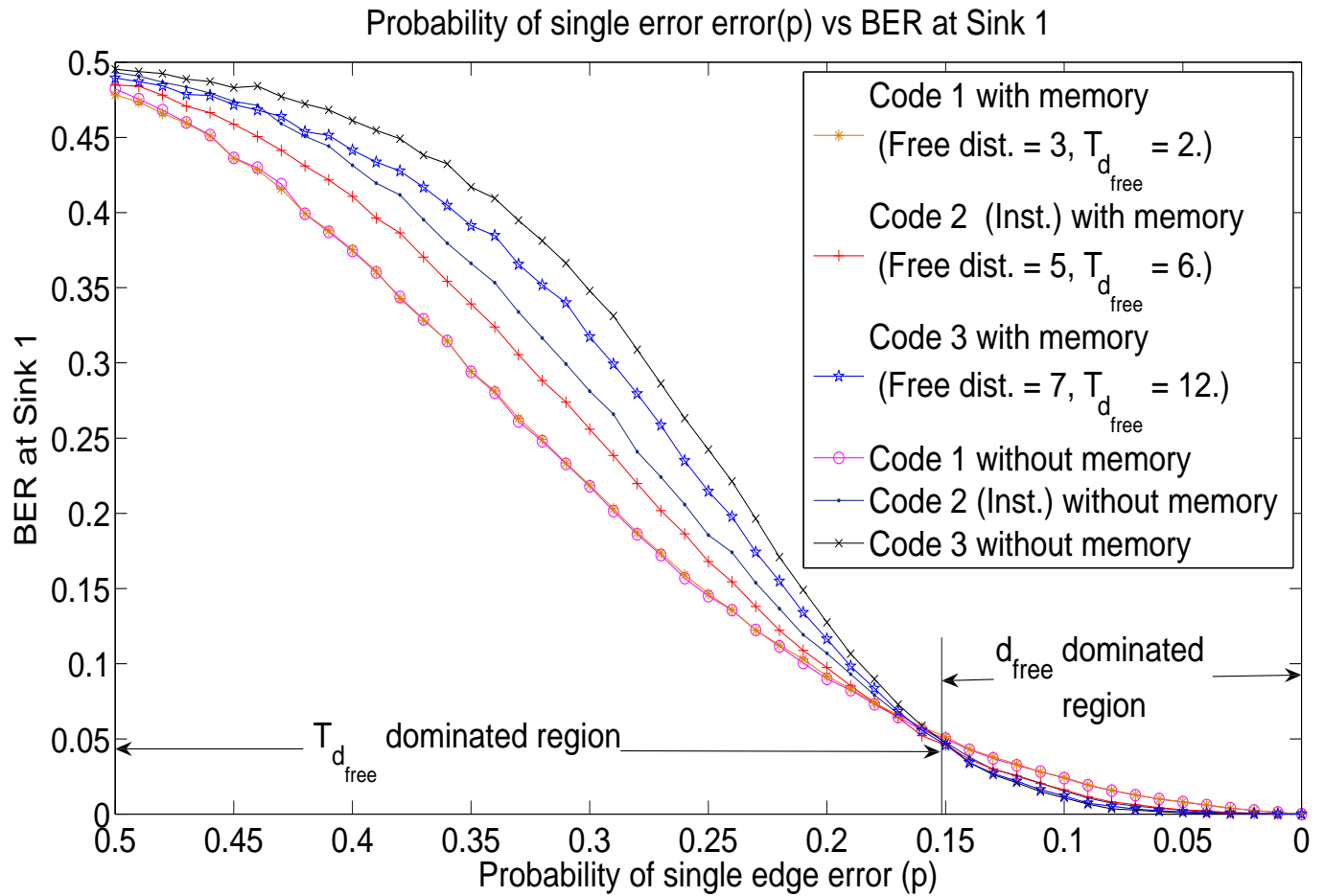}
\caption{BER (with and without memory) at Sink 1}	
\label{fig:BERsink1mem}	
\hrule
\end{figure*}

\begin{figure*}
\centering
\includegraphics[totalheight=4.90in,width=7.1in]{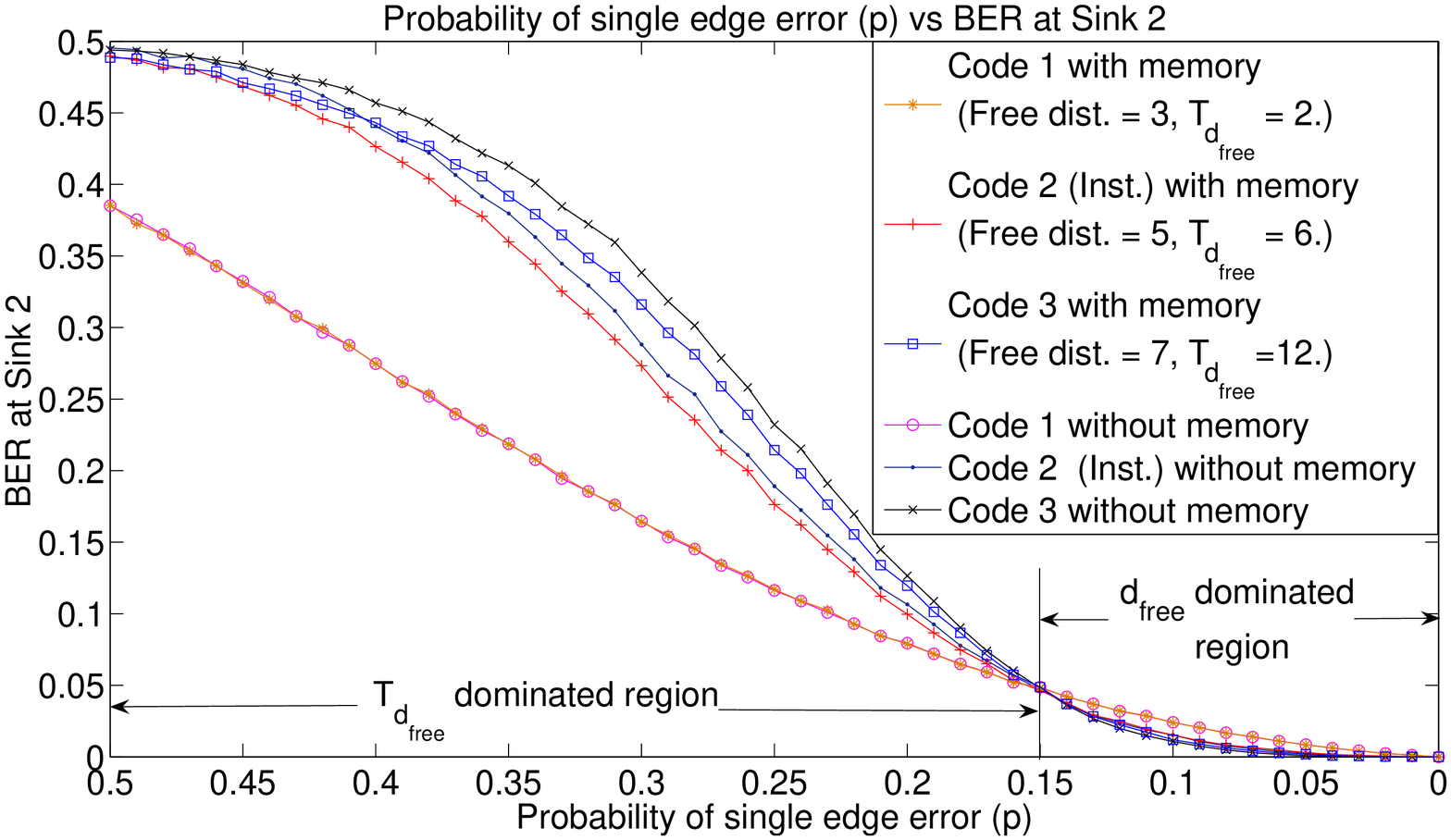}
\caption{BER (with and without memory) at Sink 2}	
\label{fig:BERsink2mem}	
\hrule
\end{figure*}

Fig. \ref{fig:BERsink1mem} and Fig. \ref{fig:BERsink2mem} illustrate the BERs for these $3$ codes for both the with-memory and memory-free case for different values of the parameter $p$ (the probability of a single edge error) of (\ref{eq:1}). Clearly the BER values fall with decreasing $p.$

The description and explanation of the regions marked `$d_{free}$ dominated region' and `$T_{d_{free}}$ dominated region' (named so according to the dominant parameter in those regions) are given in \cite{PrR}. In the following discussion, we concentrate on the comparison between the performance of every code in the memory-free and the with-memory case. Towards that end, we recall from Proposition \ref{minweighttime} that both the Hamming weight of error events and the separation between any two consecutive error events are important to correct them. 

\subsubsection*{Performance improvement of CNECCs with memory at the intermediate nodes}
\begin{enumerate}
\item With respect to codes ${\cal C}_2$ and ${\cal C}_3$, we see that there is an improvement in performance when memory is used at the intermediate nodes. This is because of the fact that the presence of memory elements in the network results in a clumping-together of error bits at the sinks. For example, assume that in the network of Fig. \ref{fig:butterflydelaymem}, an error occurs in edge $s\rightarrow v_1$ at time instant $t_1.$ We consider the situation at Sink 2. In the memory-free case, the effect of this error is felt at different time instants at the two incoming edges of Sink 2, at $t_1+1$ and at $t_1+4$. However, with memory elements at the intermediate nodes, the effects of the edge error now occur at the same time instant ($t_1+4$) in both the incoming edges of Sink 2. The effect of such errors cumulatively result in more error events (with less Hamming weights each) in the memory-free case (because of the distribution of errors) and less error events (with comparatively more Hamming weights each) in the with-memory case (as a result of clumped errors). However, because  Codes ${\cal C}_2$ and ${\cal C}_3$ have enough free distance, the number of such error events is what dominates the performance. Therefore Codes ${\cal C}_2$ and ${\cal C}_3$ correct more errors in the with-memory case. The same effect may be observed at Sink 1 also.
\item With respect to the code ${\cal C}_1$, there is no observable change in performance between the memory-free and with-memory cases. We note that the same effect is observed with the errors as in the previous case. But because of $T_{d_{free}}({\cal C}_1)$ being less (only $2$), the clumping together of error bits does not benefit much. Therefore there is no significant improvement in performance. 
\item There is no significant difference in the performance of any code between the memory-free and the with-memory case in the `$d_{free}$ dominated region.' This is because of the fact that the errors that occur in the network are already sparse.
\end{enumerate}

\section*{Acknowledgment} This work was supported  partly by the DRDO-IISc program on Advanced Research in Mathematical Engineering through a research grant to B.~S.~Rajan.

\end{document}